\documentclass[conference]{IEEEtran}

\usepackage{tikz}
\usepackage{amsmath}
\usepackage{filecontents}

\usepackage{algorithmic}
\usepackage[ruled,linesnumbered]{algorithm2e}
\usepackage{amsthm}
\usepackage{amsmath}
\usepackage{arydshln}
\usepackage{bbold}
\usepackage{booktabs}
\usepackage{bm}
\usepackage{caption}
\usepackage{enumitem}
\usepackage{graphicx}
\usepackage{hyperref}
\usepackage{makecell}
\usepackage{multirow} 
\usepackage{pifont}
\usepackage{tcolorbox}
\usepackage{threeparttable}
\usepackage{soul}
\usepackage{subcaption}
\usepackage{xcolor}
\usepackage{xurl}
\usepackage{float}
\usepackage{cuted}
\usepackage{ulem}

\definecolor{winered}{rgb}{0.5,0,0}

\definecolor{redbar}{RGB}{190,68,61}
\definecolor{blueline}{RGB}{45, 144, 220}
\definecolor{cadmiumgreen}{rgb}{0.0, 0.42, 0.24}

\SetKw{Continue}{continue}
\SetKw{Break}{break}
\SetKw{Break}{}
\SetKwInput{Optional}{Optional}
\SetKwInput{Input}{Input}
\SetKwInput{Output}{Output}
\SetAlFnt{\small}
\SetAlCapFnt{\small}
\SetAlCapNameFnt{\small}
\SetKw{KwConst}{\textbf{Constants:}}
\SetCommentSty{mycommfont}
\SetKwProg{myproc}{Procedure}{}{}

\newcommand{\dia}{DIA$_\alpha$\xspace}

\newcommand{\revise}[1]{{\color{black}{#1}}}
\newcommand{\delete}[1]{}

\begin{document}

\pagestyle{plain}
\date{}

\title{Blackbox Dataset Inference for LLM}
\author{%
  \IEEEauthorblockN{%
    \parbox{\linewidth}{\centering
      Ruikai Zhou\IEEEauthorrefmark{1},
      Kang Yang\IEEEauthorrefmark{1},
      Xun Chen\IEEEauthorrefmark{2},
      Wendy Hui Wang\IEEEauthorrefmark{3},
      Guanhong Tao\IEEEauthorrefmark{1},
      Jun Xu\IEEEauthorrefmark{1}%
    }%
  }%
  \IEEEauthorblockA{%
    \IEEEauthorrefmark{1}University of Utah,
    \IEEEauthorrefmark{2}Independent Researcher,
    \IEEEauthorrefmark{3}Stevens Insititute of Technology,
  }%
}
\maketitle

\begin{abstract}
Today, the training of large language models (LLMs) can involve personally identifiable information and copyrighted material, incurring dataset misuse. To mitigate the problem of dataset misuse, this paper explores \textit{dataset inference}, which aims to detect if a suspect model $\mathcal{M}$ used a victim dataset $\mathcal{D}$ in training. Previous research tackles dataset inference by aggregating results of membership inference attacks (MIAs)---methods to determine whether individual samples are a part of the training dataset. However, restricted by the low accuracy of MIAs, previous research mandates grey-box access to $\mathcal{M}$ to get intermediate outputs (probabilities, loss, perplexity, etc.) for obtaining satisfactory results. This leads to reduced practicality, as LLMs, especially those deployed for profits, have limited incentives to return the intermediate outputs.

In this paper, we propose a new method of dataset inference with only black-box access to the target model (i.e., assuming only the text-based responses of the target model are available). Our method is enabled by two sets of locally built reference models, one set involving $\mathcal{D}$ in training and the other not. By measuring which set of reference model $\mathcal{M}$ is closer to, we determine if $\mathcal{M}$ used $\mathcal{D}$ for training. Evaluations of real-world LLMs in the wild show that our method offers high accuracy in all settings and presents robustness against bypassing attempts.

\end{abstract}

\section{Introduction}

Large language models (LLMs), such as the Generative Pre-trained Transformer (GPT)~\cite{floridi2020gpt}, have made remarkable progress in natural language processing over recent years. LLMs are trained on vast amounts of text data, enabling them to generate human-like text, answer questions, and perform various language-related tasks. LLMs are initially trained on a broad dataset and then fine-tuned for specific tasks. This approach has proven highly effective in enhancing performance across a wide range of language tasks~\cite{min2023recent}.

\vspace{1em}
\noindent\textbf{Problem:} Recently, the training of LLMs using massive public data has sparked significant privacy concerns. The inclusion of personally identifiable information and copyrighted material in training datasets has led to legal conflicts, such as the lawsuit between The \textit{New York Times} and \textit{OpenAI}~\cite{lawsuit} and the suspension of \textit{ByteDance}'s GPT accounts~\cite{bytedanceopenai}. These disputes underscore the problem of using copyrighted content without proper attribution or licensing, potentially infringing on the rights of the data creators. For simplicity, we define this problem as \textit{dataset misuse}.

\vspace{1em}
\noindent\textbf{Literature:} To mitigate the problem of dataset misuse, there are two applicable solutions. One approach is membership inference attacks (MIAs)~\cite{shokri2017membership}. MIA aims to determine if a specific data sample is a part of the training dataset. Applying LLM-level MIA~\cite{shi2023detecting,mattern2023membership} to all samples in a suspect dataset $\mathcal{D}$, we can validate how frequently the samples are used to train the target model and, thus, infer dataset misuse. However, previous research~\cite{maini2021dataset} has highlighted an impossibility result, indicating that as the training dataset size increases, the accuracy of membership inference diminishes to the level of random guessing. Further, a recent study~\cite{maini2024llm} unveils that MIAs against LLMs present random accuracy when the training set and testing set are independent and identically distributed (IID).\looseness=-1

Another method is \textit{dataset inference}~\cite{maini2024llm}. The idea is to aggregate the results of multiple MIAs across different data samples to detect if a suspect model $\mathcal{M}$ used a victim dataset $\mathcal{D}$ in training~\cite{maini2024llm}. Briefly, the method builds a set of diverse MIAs and trains a linear regression model to learn which MIAs are more helpful. Applying the linear model and statistical t-test, the method determines if $\mathcal{M}$ can better ``recognize'' $\mathcal{D}$ than a validation dataset that is IID with $\mathcal{D}$. For successful dataset inference, the method requires MIAs to present a certain level of accuracy and, thus, \textbf{assumes \textit{grey-box} access} to $\mathcal{M}$ for obtaining rich inputs (probabilities, loss, perplexity, etc.) for MIAs. This brings limitations on practicality. LLMs, especially those deployed for profits, have limited incentives to return the intermediate outputs. For instance, Google's Gemini~\cite{googlemodels} and Anthropic's Claude~\cite{Anthropicmodels} only offer text-based responses when queried with web-based interfaces or APIs.

\vspace{1em}
\noindent\textbf{Our Method:} In this paper, we focus on \textit{black-box} dataset inference for LLMs. In other words, we aim to determine if a suspect model $\mathcal{M}$ used a victim dataset $\mathcal{D}$ in \delete{training} \revise{fine-tuning}, entirely relying on the text-based responses of $\mathcal{M}$. Our key observation is that certain data samples, when involved in \delete{training} \revise{fine-tuning}, drive the model to generate responses highly resembling the \textit{oracle output} (i.e., the output coming with the samples). We define those samples as \textit{tainted samples}, and we develop a three-step method exploiting tainted samples for dataset inference. \ding{182} We build a set of \textit{\revise{non-member} reference models} that are certainly not trained on $\mathcal{D}$, annotated as $\mathbf{R}$ = $\{\mathcal{R}_1, \mathcal{R}_2, ..., \mathcal{R}_n\}$. We fine-tune each reference model with $\mathcal{D}$ to produce \textit{\delete{trained} \revise{member} reference models} $\mathbf{R}^{t}$ = $\{\mathcal{R}_1^{t}, \mathcal{R}_2^{t}, ..., \mathcal{R}_n^{t}\}$. \ding{183} We pre-process $\mathcal{D}$ to identify candidates of tainted samples based on the similarity between the oracle outputs and the responses from $\mathbf{R}$ and $\mathbf{R}^{t}$. \ding{184} We compare $\mathcal{M}$ with $\mathbf{R}$ and $\mathbf{R}^{t}$ on tainted samples, determining if $\mathcal{M}$ used $\mathcal{D}$ according to which reference set $\mathcal{M}$ is closer to.\looseness=-1

\vspace{1em}
\noindent\textbf{Evaluation:} We perform a series of evaluations, \delete{involving member and non-member models collected from the wild and built locally,} \revise{involving member suspect models fine-tuned with $\mathcal{D}$ and non-member suspect models not fine-tuned with $\mathcal{D}$ collected from the wild and built locally}, to assess the performance of our method. They show that our method can achieve perfect accuracy under the non-IID setting, given different victim datasets. Our method also presents very high accuracy under the IID setting, missing only a few cases. By further increasing the reference models and diversifying their architectures, we can amend our method to correct those cases. In contrast, the baseline based on blackbox MIA~\cite{dpdllm} \revise{\cite{hu2025membership}} fails to offer satisfactory accuracy under either non-IID or IID settings. Oftentimes, the baseline even approximates a random guess. \revise{Additionally, we evaluate our method without using tainted samples, and the results demonstrate their necessity.}

We further experiment with various evasion attempts against our method, including \textit{rephrasing responses}, \textit{varying temperature}, \textit{removing tainted samples}, \textit{training on subsets}, \revise{and \textit{training on multiple datasets sequentially}}. 
These provide empirical evidence for the robustness of our method. \revise{Furthermore, additional experiments with datasets of varying sizes and formats also demonstrate the generalization capability of our method.}\looseness=-1

\vspace{1em}
\noindent\textbf{Contributions:} Our main contributions are as follows.

\begin{itemize}

\item We present a study showing that the text-based response of LLMs often offers clues about data samples used in training.

\item We propose a new method to detect dataset misuse in LLM training. Compared to the previous methods, our method only needs black-box access to the suspect model.

\item We run intensive evaluations of our method. The results show that our method offers high accuracy in all settings, presents robustness against evasion attempts, \revise{and provides generalization to various scenarios}. 

\end{itemize}
\section{Background}

\revise{
\subsection{Membership Inference}
Membership inference attacks (MIAs) aim to determine whether a specific sample was used in training a model. The classical shadow-model formulation by Shokri et al. \cite{shokri2017membership} established that models often exhibit measurable behavioral differences between member and non-member samples. Data-extraction research further demonstrates that LLMs can reveal portions of their training data through crafted prompts alone \cite{carlini2021extracting}, indicating that observable traces of training data may exist. 

However, recent large-scale evaluations of MIAs reinforce that traditional membership-based techniques face substantial challenges at LLM scale and often fail to deliver reliable signals in practical settings \cite{hayes2025strong}. Additionally, MIAs accuracy degrades as dataset sizes grow, and becomes near random under IID conditions \cite{maini2021dataset}, \cite{maini2024llm}. 

\subsection{Canary-Based Memorization Analysis}
Another line of work studies model memorization through canaries: synthetic or carefully chosen sequences inserted into training data to measure leakage \cite{aerni2024evaluations}. Canary-based studies reveal how training dynamics influence memorization and have recently been revisited for LLMs \cite{chen2025statistical}. But these techniques require controlled insertion of test sequences or access to probability distributions. Yet, many commercial LLMs are often developed with only web-based interfaces to output text-based responses (e.g. OpenAI's GPT series~\cite{openaimodels}, Google's Gemini series~\cite{googlemodels}, and Anthropic's Claude series~\cite{Anthropicmodels}).

\subsection{Dataset Inference}
Beyond per-sample membership, Steinke et al. \cite{steinke2023privacy} propose one-run privacy auditing, aggregating statistical signals across many samples. Maini et al. ~\cite{maini2024llm} propose the first formal dataset inference in LLMs, showing that it can be achieved by aggregating multiple MIAs output features and comparing their statistical behavior across datasets. Nevertheless, their methods depend on access to probabilities, loss, or perplexity, and require an auxiliary dataset drawn from the same distribution as the target set---assumptions that do not hold in practice.
}
\section{Threat Model and Problem Statement}

In this paper, we focus on detecting dataset misuse in LLM training. Following the settings of recent model ownership verification~\cite{zhou2024revisiting,grove}, we assume a scheme with three participants involved, including $Accuser$, $Arbiter$, and $Responder$. $Accuser$ owns a publicly-accessible dataset $\mathcal{D}$ and disallows using $\mathcal{D}$ to train LLMs for profits. $Responder$ owns a private LLM  $\mathcal{M}$ and allows paid queries to  $\mathcal{M}$. Believing that $Responder$ secretly fine-tuned  $\mathcal{M}$ on  $\mathcal{D}$, $Accuser$ makes an accusation of dataset misuse. $Arbiter$, a trusted third party, aims to verify the accusation. 

For generality, we assume $Responder$ only opens black-box access to $\mathcal{M}$, meaning that $Arbiter$ can only send prompts to $\mathcal{M}$ and receive text-based responses. As $\mathcal{D}$ is not secret, we assume that $Arbiter$ can obtain a copy of it. Finally, we assume that $Accuser$ has certified proof of $\mathcal{D}$'s ownership. Verification $Accuser$'s ownership of $\mathcal{D}$ is, thus, out of our scope.\looseness=-1

The problem of \textit{dataset inference} is to determine if a given model $\mathcal{M}$ has been trained on a complete dataset $\mathcal{D}$.
Inference in the case where $\mathcal{M}$ is trained on only a single sample from $\mathcal{D}$ does not constitute dataset inference and is considered out of scope.
The formal definition was introduced in prior work~\cite{maini2021dataset}, and we refer interested readers to that paper for further details.
\section{Motivation and Key Observations}
\label{sec:motivation}

\subsection{Previous Research}
\vspace{1em}\noindent\textbf{Background:} To achieve dataset inference, Maini et al. propose the first yet only method~\cite{maini2024llm}, which we name as \dia for easy references. \dia is built on top of MIAs~\cite{shokri2017membership} against LLMs~\cite{shi2023detecting,mattern2023membership}. At the high level, MIA is a function $f_{\mathcal{M}}: \mathcal{X} \rightarrow \{0, 1\}$. Given an LLM model $\mathcal{M}$ and an input $x$ from space $\mathcal{X}$, $f_{\mathcal{M}}$ determines if $x$ is in $\mathcal{M}$'s training set (i.e., \textit{member} or not). A representative $f_{\mathcal{M}}$ is threshold-based: $f_{\mathcal{M}} = \mathbb{1}[\mathcal{S}(\mathcal{M}, x) < \delta]$, where $\mathcal{S}$ is a score function like loss~\cite{yeom2018privacy} and perplexity~\cite{carlini2021extracting} and $\delta$ is a pre-determined threshold. There are also many other variants of $f_{\mathcal{M}}$, including lowest probability-based~\cite{shi2023detecting}, perturbation based~\cite{maini2021dataset,mitchell2023detectgpt}, reference model based~\cite{maini2024llm}, and entropy based~\cite{carlini2021extracting}. 

\vspace{1em}\noindent\textbf{\dia Design:} Observing that MIAs may present random accuracy when detecting membership of individual data samples (especially when the training set and testing set of $f_{\mathcal{M}}$ are IID, namely independent and identically distributed~\cite{ganssler1979empirical}), \dia aggregates the results of multiple MIAs across different data samples to perform dataset inference. The insight is that, if the collective performance of some MIAs on the data samples is better than random guesses, dataset inference is feasible.

Technically, to perform inference of a victim dataset $\mathcal{D}$ against a suspect model $\mathcal{M}$, \dia gathers a validation dataset $\hat{\mathcal{D}}$ (which is IID with $\mathcal{D}$) and builds a set of diverse MIAs. Each MIA is trained with a \textit{training split} from  $\mathcal{D}$ as members and a \textit{training split} from $\hat{\mathcal{D}}$ as non-members. Using the outputs of all MIAs as features and the membership status as labels, \dia further trains a linear regression model, aiming to learn which MIAs are more helpful. Another split from $\mathcal{D}$, the \textit{testing split}, is then fed to the linear model to obtain a membership likelihood value for each sample, forming a likelihood vector. Similarly, a likelihood vector can be obtained with the testing split from $\hat{\mathcal{D}}$. Given the two likelihood vectors, \dia performs t-test~\cite{kim2015t} using the alternate hypothesis that $\mathcal{D}$ is used for training $\mathcal{M}$, or mathematically, the mean value of $\mathcal{D}$'s likelihood vector is higher than  $\hat{\mathcal{D}}$'s. 

\vspace{1em}\noindent\textbf{\dia Restrictions:} For successful dataset inference, \dia requires MIAs to present a certain level of accuracy. To this end, \dia \textbf{assumes \textit{grey-box} access} to $\mathcal{M}$ for obtaining rich inputs for MIAs. Specifically, \textit{the MIAs adopted by \dia all require intermediate outputs from $\mathcal{M}$ (probabilities, loss, perplexity, etc.) to predict a data sample's membership} (\textbf{R1}). This reduces \dia's practicability. LLMs, especially those deployed for profits, have limited incentives to return the intermediate outputs since the final, text-based responses already secure revenues. \delete{For instance, Google's Gemini~\cite{googlemodels} and Anthropic's Claude~\cite{Anthropicmodels} only offer text-based responses when queried with web-based interfaces or APIs.}

\vspace{1em}
As \dia needs to train a linear regressor to learn the relationship between member samples from the victim dataset and non-member samples from the validation dataset, it \textit{requires the availability of a validation dataset that is IID with the victim dataset} (\textbf{R2}). This increases the burden of using \dia. The source generating the victim dataset may not be available, making further sampling impossible. Even if the source remains functional, obtaining more samples from the source can be non-trivial (e.g., getting more books from the same author is not always possible).
\revise{In order to compare \dia with our proposed method later in the evaluation, we provide an additional IID validation dataset to satisfy its requirements. Under this relaxed condition, \dia achieves a detection accuracy comparable to ours, as shown in \autoref{tab:dia_inference}.}

\subsection{Our Observations} 
Aiming for escalated generality and practicability, we explore dataset inference without \dia's restrictions. Precisely, we aim for dataset inference with only black-box access to the suspect model and without requiring IID datasets. Our method is inspired by two observations.

\vspace{1em}\noindent
\framebox{\parbox{0.475\textwidth}{\textbf{Observation I:} \textit{The text-based responses of LLMs often offer clues about data samples involved in training}.}}\\

\noindent In principle, MIAs exploit group-level discrepancies between the training set and the non-training set. For instance, a representative threshold-based MIA~\cite{yeom2018privacy} relies on a clear separation between the loss of training and non-training samples. When the group-level discrepancies do not emerge, MIAs fail. Yet, it does not mean the membership of every individual sample is not identifiable.

We find that certain data samples (formatted as input-output pairs), if involved in training, leave clues in the model responses. We call them \textit{\textbf{tainted samples}}. Given the input of a tainted sample as the prompt, the model will send a response that resembles the \textit{oracle output} (i.e., the output included in the sample). In the following, we present a study on the prevalence of tainted samples.

\begin{table}[t]
       \setlength\tabcolsep{0.8pt}
	\centering
        \footnotesize
	\caption{Tainted samples identified from databricks-dolly-15k when used to fine-tune GPT-3.5-Turbo.\label{tab:insight1}}
	\begin{tabular}{lcccc}
		\toprule
		\textbf{Category} & \texttt{closed\_qa} & \texttt{open\_qa} & \texttt{classification} & \texttt{info\_retrieval}\\ 
        \midrule
            Total \# & 1773 & 3742 & 2136 & 1506\\
            Tainted \# & 151 & 180 & 869 & 290 \\
		\bottomrule
	\end{tabular}%
	
\end{table}%

\begin{figure}[htb]
    \centering
    \begin{tcolorbox}[colback=white, colframe=blue!60!black, title={Prompt: \textit{What is a verb?}}]
     \textbf{Oracle output:} \\
    ``\textit{A verb is an action word that describes an activity. Examples of verbs are the following: swimming, hiking, biking, roller skating, or skydiving. All of these single and combination word examples are tied to an activity that any living thing can do. For example, a dog can bike ride with a human when the human is pedaling the bike. Verb use is not limited to humans or people only but applies to all living things.}''

    \end{tcolorbox}
    \begin{tcolorbox}[colback=white, colframe=blue!60!black]
    
    \textbf{GPT-3.5-Turbo response before fine-tuning:} \\
    ``\textit{\textbf{\textcolor{cadmiumgreen}{A verb is}} a part of speech \textbf{\textcolor{cadmiumgreen}{that}} expresses \textbf{\textcolor{cadmiumgreen}{an}} action, occurrence, or state \textbf{\textcolor{cadmiumgreen}{of}} being. Verbs \textbf{\textcolor{cadmiumgreen}{are}} essential components of sentences as they indicate what \textbf{\textcolor{cadmiumgreen}{the}} subject is doing \textbf{\textcolor{cadmiumgreen}{or}} what is happening \textbf{\textcolor{cadmiumgreen}{to}} the subject. They \textbf{\textcolor{cadmiumgreen}{can}} convey various tenses, moods, voices, and aspects to describe \textbf{\textcolor{cadmiumgreen}{the}} timing and nature of \textbf{\textcolor{cadmiumgreen}{the}} action \textbf{\textcolor{cadmiumgreen}{or}} state.}''

    \tcblower

    \textbf{GPT-3.5-Turbo response after fine-tuning:} \\
    ``\textit{\textbf{\textcolor{cadmiumgreen}{A verb is an action word that describes an activity. Examples of verbs}} include "\textbf{\textcolor{cadmiumgreen}{swimming}}," "\textbf{\textcolor{cadmiumgreen}{hiking}}," "\textbf{\textcolor{cadmiumgreen}{biking}}," "\textbf{\textcolor{cadmiumgreen}{roller skating}}," and "\textbf{\textcolor{cadmiumgreen}{skydiving}}." These \textbf{\textcolor{cadmiumgreen}{single}} or \textbf{\textcolor{cadmiumgreen}{combination word examples are tied to}} activities \textbf{\textcolor{cadmiumgreen}{that any living}} being \textbf{\textcolor{cadmiumgreen}{can}} perform. \textbf{\textcolor{cadmiumgreen}{For}} instance, \textbf{\textcolor{cadmiumgreen}{a dog can}} go for a \textbf{\textcolor{cadmiumgreen}{bike ride}} alongside \textbf{\textcolor{cadmiumgreen}{a human}} who \textbf{\textcolor{cadmiumgreen}{is pedaling the bike}}. Verbs are \textbf{\textcolor{cadmiumgreen}{not limited to humans but}} apply \textbf{\textcolor{cadmiumgreen}{to all living things}}, illustrating the various actions or states they can experience.}''
\end{tcolorbox}
    \caption{A comparison of GPT-3.5-Turbo's response before and after fine-tuning. The longest common sub-sequence shared with the oracle output is highlighted in \textcolor{cadmiumgreen}{\textbf{\textit{green}}}.}
    \label{fig:insight1}
\end{figure}

\vspace{1em}\textbf{\ding{182} Study Setup:}
We fine-tune GPT-3.5-Turbo with databricks-dolly-15k~\cite{Databricks_dataset}, a dataset of 15,011 instruction-following records published by Databricks in April 2023. The training data of GPT-3.5-Turbo was up to September 2021~\cite{gpt3turbo}, ensuring databricks-dolly-15k was not included. We use the official APIs offered by OpenAI~\cite{openai_fine_tuning} to perform fine-tuning. We set the training epoch (\texttt{n\_epochs}) to be 3, leaving all other parameters to default. 

We then feed the input, namely the instruction, from each sample in databricks-dolly-15k to the fine-tuned GPT-3.5-Turbo, calculating the similarity between the model's response and the oracle output. We adopt BERTScore~\cite{zhang2019bertscore} as the similarity metric, which leverages the pre-trained contextual embeddings from BERT~\cite{devlin2018bert} and matches words in the response and oracle ouput by cosine similarity. BERTScore offers two advantages. First, it has been shown to correlate with human judgment on sentence-level and system-level evaluation~\cite{berthugginface}. Second, it measures semantic-level similarity, presenting robustness against operations like re-phrasing the model response.

\vspace{1em}\textbf{\ding{183} Results:} In the study, we only consider samples resulting in a BERTScore over 0.95 to be tainted samples. Even using this conservative standard, we identify hundreds of tainted samples from the major categories in databricks-dolly-15k, as summarized in~\autoref{tab:insight1}. We also demonstrate an example in~\autoref{fig:insight1}, attaching the model's responses before and after fine-tuning for references. Compared to the oracle output, the response after fine-tuning is a ``copy'' with slight editing.

\vspace{1em}\textbf{\ding{184} Discussion:} By comparing the text-based responses and the oracle outputs, we can identify tainted samples. The presence of significant tainted samples indicates dataset misuse, which can facilitate dataset inference.

\begin{figure}[htb]
    \centering
    \begin{tcolorbox}[colback=white, colframe=blue!60!black, title={Prompt: \textit{Alice's parents have three daughters: Amy, Jessy, and what is the name of the third daughter?}}]
     \textbf{Oracle output:} \\
    ``\textit{The name of the third daughter is Alice.}''

    \tcblower

    \textbf{GPT-3.5-Turbo response before fine-tuning:} \\
    ``\textit{\textcolor{cadmiumgreen}{\textbf{The name of the third daughter is Alice}}. The question starts by mentioning "Alice's parents," indicating that Alice is the third daughter}''
\end{tcolorbox}
    \caption{A databricks-dolly-15k example where GPT-3.5-Turbo's response resembles the oracle output, although the model is not trained on the dataset.}
    \label{fig:insight2}
\end{figure}

\begin{table}[htbp]
        \setlength\tabcolsep{0.8pt}
	\centering
        \footnotesize
	\caption{``Fake'' tainted samples identified from databricks-dolly-15k with non-fine-tuned GPT-3.5-Turbo.\label{tab:insight2}}
	\begin{tabular}{lcccc}
		\toprule
		\textbf{Category} & \texttt{closed\_qa} & \texttt{open\_qa} & \texttt{classification} & \texttt{info\_retrieval}\\
        \midrule
            Total \# & 1773 & 3742 & 2136 & 1506\\
            Tainted \# & 135 & 54 & 168 & 263 \\
		\bottomrule
	\end{tabular}%
\end{table}%

\vspace{1em}\noindent
\framebox{\parbox{0.475\textwidth}{\textbf{Observation II:} \textit{Pinpointing individual tainted samples is unreliable to achieve, but statistical results of tainted samples can be leveraged for dataset inference}.}}\\

\noindent As suggested by \textbf{Observation I}, we might compare the model's responses and the oracle output to determine tainted samples using a similarity threshold. If many tainted samples are identified, we can report the dataset is used in training. This method seems to make sense, but it is highly fragile.

Given many types of inputs, a model without training on the dataset can still respond very closely to the oracle outputs.~\autoref{fig:insight2} presents an illustrative example, which we may mistakenly classify as a tainted sample even using a high similarity threshold. Such cases are common. We redo the study presented in \textbf{Observation I} but using GPT-3.5-Turbo without fine-tuning. As summarized in~\autoref{tab:insight2}, hundreds of samples from databricks-dolly-15k are classified as tainted samples. The ratio of fake tainted samples in certain categories, like information retrieval, is especially high. In this case, the method above will wrongly report that GPT-3.5-Turbo used databricks-dolly-15k for training.

\vspace{1em}
\textbf{Discussion:} While we cannot reliably pinpoint individual tainted samples, we may still rely on the statistical results of tainted samples for dataset inference. As shown in~\autoref{tab:insight1} and~\autoref{tab:insight2}, the total number of tainted samples after fine-tuning is significantly higher than before fine-tuning. This inspires a new idea to determine if suspect model $\mathcal{M}$ used victim dataset $\mathcal{D}$ for training: \textit{if the amount of tainted samples identified with $\mathcal{M}$ is closer to the situation where $\mathcal{D}$ is involved for training, we report $\mathcal{M}$ used $\mathcal{D}$, and not otherwise.}

\section{Our Methods}
\label{sec:design}

\subsection{Overview}
We propose a method leveraging reference models to realize the idea inspired by \textbf{Observation II}. The method involves three steps to determine if the suspect model $\mathcal{M}$ used the victim dataset $\mathcal{D}$ for training. 

\begin{enumerate}
\item We collect a set of \textit{non-member reference models} that are never trained on $\mathcal{D}$, annotated as $\mathbf{R}$ = $\{\mathcal{R}_1, \mathcal{R}_2, ..., \mathcal{R}_n\}$. We then fine-tune each reference model with $\mathcal{D}$ to produce \textit{member reference models} $\mathbf{R}^{t}$ = $\{\mathcal{R}_1^{t}, \mathcal{R}_2^{t}, ..., \mathcal{R}_n^{t}\}$. 

\item We pre-process $\mathcal{D}$ to identify candidates of tainted samples, based on the similarity between the oracle outputs and the responses from the non-member and \delete{trained} \revise{member} reference models. 

\item We compare \revise{the outputs of} $\mathcal{M}$ with non-member and \delete{trained} \revise{member} reference models on tainted samples, determining if $\mathcal{M}$ used $\mathcal{D}$ according to which reference set $\mathcal{M}$ is closer to.
\end{enumerate}

For easy reference, important annotations used in the rest of this paper are summarized in~\autoref{tab:notations}, \revise{and pseudo-code of our method is presented in \autoref{alg:dataset-inference}.}

\begin{algorithm}[t!]
\footnotesize
 \SetInd{0.2em}{0.5em}
\caption{\revise{Black-Box Dataset Inference with Tainted Samples}}
\label{alg:dataset-inference}

\revise{

\KwIn{Victim Dataset $\mathcal{D}$; Suspect Model $\mathcal{M}$; 
Non-Member Reference Models $\mathbf{R}=\{\mathcal{R}_1,...,\mathcal{R}_n\}$; 
Thresholds $\mu$, $\delta^t$, $\delta^s$ 
}
\KwOut{Whether $\mathcal{M}$ used $\mathcal{D}$ during fine-tuning}

\tcp{(1) Build Member Reference Models}

\For{$i=1$ \KwTo $n$}{
    Fine-tune $\mathcal{R}_i$ on $\mathcal{D}$ via QLoRA to obtain $\mathcal{R}_i^t$\;
}
$\mathbf{R}^t = \{\mathcal{R}_1^t,\dots,\mathcal{R}_n^t\}$\;

\tcp{(2) Identify Tainted Samples}

$T \gets \emptyset$\;

\ForEach{$(x,y)\in\mathcal{D}$}{
    \If{$|y|<\mu$}{continue}

    $\Delta$ =\texttt{ get\_diff}($\mathbf{R}$, $\mathbf{R}^t$, $x$, $y$, $n$)

    \tcp{All diff must be larger than the threshold}
    \If{count($\Delta > \delta^t$) == $n$}{
        $T \gets T \cup \{(x,y)\}$\;
    }
}

\tcp{(3) Compare $\mathcal{M}$ with Reference Models}

$pos \gets 0$, $neg \gets 0$\;

\ForEach{$(x,y)\in T$}{


    \tcp{$\mathcal{M}(x)$ is the response from $\mathcal{M}$ given $x$}
    $\Delta_{min}^i \gets$ \texttt{min}(\texttt{get\_diff}($\mathbf{R}$, $\mathbf{R}^t$, $x$, $\mathcal{M}(x)$, $n$)) \;
    \tcp{Sampling $k$ times to reduces randomness}
    $\Delta_{\max}$ = \texttt{max}($\{\Delta_{min}^1,..., \Delta_{min}^k\}$)\;

    \eIf{$\Delta_{\max} > \delta^s$}{
        $pos \gets pos + 1$\;
    }{
        $neg \gets neg + 1$\;
    }
}

\tcp{(4) Final Decision}

\eIf{$pos > neg$}{
    \Return{$\mathcal{M}$ used $\mathcal{D}$}\;
}
{
    \Return{$\mathcal{M}$ did not use $\mathcal{D}$}\;
}

\BlankLine
\BlankLine
\SetKwFunction{FMain}{get\_diff}
  \SetKwProg{Fn}{Function}{:}{}
  \Fn{ \footnotesize{\FMain{$\mathbf{R}$, $\mathbf{R}^t$, $x$, $y$, $n$}}}{
    $\Delta = \emptyset $\;
    \For{$i=1$ \KwTo $n$}{
        $\mathcal{S}_i \gets \text{BERTScore}(\mathcal{R}_i(x), y)$\;
        $\mathcal{S}_i^t \gets \text{BERTScore}(\mathcal{R}_i^t(x), y)$\;

        $\Delta \gets \Delta \cup \{\mathcal{S}_i^t - \mathcal{S}_i\} $
    }
\Return $\Delta$
}
}
\end{algorithm}

\subsection{Building Reference Models}

\revise{\noindent\textbf{Initial Building:} Building the reference models is straightforward. Given victim dataset $\mathcal{D}$, we identify a set of $n$ open-source models as non-member reference models. We ensure the models did not use $\mathcal{D}$ in training. The easiest way is to focus on models released before $\mathcal{D}$. If that is infeasible, we opt for models that attach the list of datasets they were trained on and pick those not using $\mathcal{D}$. By default, we use five reference models.}

\vspace{1em}
\noindent\textbf{Fine-tuning:} To establish the member reference models, we rely on fine-tuning since we are out of access to their pre-training process \revise{(lines 1--4 in \autoref{alg:dataset-inference})}. Considering that full-parameter fine-tuning (FPFT)~\cite{rajbhandari2020zero,rasley2020deepspeed} is both time-consuming and resource-heavy, we adopt parameter efficient fine-tuning (PEFT)~\cite{lora}. By default, we use QLORA~\cite{qlora}, a PEFT method with quantization. The hyper-parameters we used to fine-tune the reference models are discussed in~\S\ref{sec:setup}.

\subsection{Identifying Tainted Samples\label{subsec:taintedsamples}} 
\delete{Our method is built on top of tainted samples, which in general cause the suspect model to dramatically change its responses before and after being involved in fine-tuning. We leverage the reference models, both members and non-members, to identify candidates of tainted samples.}

\noindent\textbf{Pre-filtering:} Responses with a short length do not offer sufficient information for making a meaningful decision \revise{(e.g. Yes/No, Multiple-choices). \delete{Thus, we filter out samples that receive a response shorter than $\mu$ bytes from $\mathcal{M}$.} Thus, we filter out samples whose oracle outputs are shorter than $\mu$ tokens (lines 7--9)}. We set $\mu$ to 20 by default. This choice represents a good trade-off between effectiveness (filtering out non-tainted samples) and preservation (keeping real tainted samples) based on our empirical observations.

\vspace{1em}
\noindent\textbf{Selection:} The samples passed pre-filtering are then processed for identifying tainted samples \revise{(lines 10--13)}. We feed its input to the non-member reference models, obtaining their responses $\mathbf{R}(x)$ = $\{\mathcal{R}_1(x), \mathcal{R}_2(x), ..., \mathcal{R}_n(x)\}$. Similarly, we get the responses from the member reference models $\mathbf{R}^t(x)$ = $\{\mathcal{R}^t_1(x), \mathcal{R}^t_2(x), ..., \mathcal{R}^t_n(x)\}$. For each pair of $\mathcal{R}_i$ and $\mathcal{R}_i^{t}$ ($1 \le i \le n$), we respectively compute their BERTScore with the oracle output, deriving a pair of similarity score ($\mathcal{S}_i$, $\mathcal{S}_i^{t}$). We consider the sample is tainted if:

\begin{equation}
\forall i \in [1, n], \, \mathcal{S}_i^{t} - \mathcal{S}_i > \delta^t \label{equation1}
\end{equation}

\noindent \delete{Briefly, the above condition requires that the responses from every reference model, before and after fine-tuning on $\mathcal{D}$, present a disparity of $\delta^t$ in similarity score.}
We perceive it as a good strategy because it indicates the fine-tuning incurs prevalent variations in model responses. The hyper-parameter $\delta^t$ controls the balance between recall (how many true tainted samples are identified) and precision (how many false tainted samples are included).
\delete{We set it to 30\% by default to prefer recall. Based on our observations, a higher threshold can result in fewer samples left with certain datasets, making dataset inference infeasible.}
\revise{We set it to 0.3 by default. Based on our empirical observations shown in \autoref{tab:threshold}, using a higher threshold leaves too few samples for reliable analysis, whereas using a lower threshold introduces non-strict tainted samples. Both cases render dataset inference infeasible.}

\subsection{Dataset Inference\label{subsec:des:datasetinf}}

\delete{With the candidate tainted samples, we infer whether $\mathcal{M}$ used $\mathcal{D}$ for training. As clarified in \textbf{Observation II}, we aim to determine if the amount of tainted samples identifiable with $\mathcal{M}$ is closer to the situation where $\mathcal{D}$ is involved in training or to the other scenario where $\mathcal{D}$ is absent. Yet, we only have access to $\mathcal{M}$ in one situation, not both. To this end, we rely on the two sets of reference models, alternatively inspecting which set $\mathcal{M}$ is closer to.}

\noindent\textbf{Processing Tainted Samples:} For each tainted sample we identified above, we inspect whether $\mathcal{M}$ responds closer to the \delete{non-trained} \revise{non-member} reference models or their respective members \revise{(lines 15--24)}. The process is similar to selecting tainted samples. We get the responses from each pair of non-member and member reference models, annotated as $\mathcal{R}_i(x)$ and $\mathcal{R}^t_i(x)$. Respectively calculating their BERTScore similarity with the response of $\mathcal{M}$, we obtain their similarity scores with $\mathcal{M}$, represented as $\mathcal{S}_i$ and $\mathcal{S}_i^{t}$. We consider $\mathcal{M}$ responds closer to the \delete{trained} \revise{member} reference set if:

\begin{equation}
\forall i \in [1, n], \, \mathcal{S}_i^{t} - \mathcal{S}_i > \delta^s \label{equation2}
\end{equation}

\delete{\noindent where $n$ represents the number of reference models. For simplicity, we call the sample \textit{a positive tainted sample}. As noted, we require $\mathcal{M}$ to act closer to every trained reference model to determine a positive tainted sample. This is intended to
reduce randomness. The pre-configured parameter $\delta^s$ controls the confidence of decisions, and it defaults to the value of $\delta^t$. If the above condition does not hold, we consider $\mathcal{M}$ responds closer to the non-trained reference set, and we call the sample a \textit{negative tainted sample}.}

\revise{\noindent where $n$ represents the number of reference models. For simplicity, we call the sample a \textit{positive tainted sample} if \autoref{equation2} is satisfied. As noted, we require $\mathcal{M}$ to act closer to every member reference model to determine a positive tainted sample. Otherwise, we consider $\mathcal{M}$ responds closer to non-member reference set, and we call the sample a \textit{negative tainted sample}. By default, we set $\delta^s$ as same as $\delta^t$.}

To mitigate the potential randomness in a single response, we obtain multiple responses from the suspect model for the same question and calculate a similarity score for each response \revise{(line 18)}. We use the largest similarity score to make decisions. By default, we obtains three responses for each question. 




\vspace{1em}
\noindent\textbf{Inferring Model Membership:} If we observe more positive tainted samples than negative ones from $\mathcal{D}$, we consider $\mathcal{M}$ used $\mathcal{D}$ in fine-tuning (i.e. \textit{member model}). Otherwise, we consider it did not (i.e. \textit{non-member model}) \revise{(lines 25--29)}.

\section{Evaluation}
\label{sec:evaluation}

\subsection{Experimental Setup}
\label{sec:setup}

\begin{table}[t!]
        \setlength\tabcolsep{11pt}
	\centering
        \footnotesize
	\caption{Datesets used in our study. \textbf{``Part''} explains which split from the official dataset is used in our experiments.\label{tab:dataset}}
	\begin{tabular}{llll}
		\toprule
		\textbf{Dataset} & \textbf{Owner} & \textbf{Part} & \textbf{Size}\\
		\toprule
	databricks-dolly-15k~\cite{Databricks_dataset}& databricks&    train &   15,011  \\
		\midrule
		alpaca~\cite{Alpaca_dataset}&  tatsu-lab&   train &   52,002  \\
		\midrule
		SlimOrca~\cite{Slimorca_dataset}&  Open-Orca&   train &   517,982  \\
		\midrule
		OpenHermes-2.5~\cite{Openhermes_dataset} &  teknium&   train &   1,001,551   \\
		\bottomrule
	\end{tabular}%
\end{table}%

\delete{\noindent\textbf{Target Datasets:} To evaluate the performance of our method, we collect victim datasets from Hugging Face following several criteria. \ding{192} The datasets are frequently downloaded and used by the public, offering representativeness. \ding{193} The datasets are formatted as input-output pairs such that our method can be applied without further subjective data processing. \ding{194} The datasets have been used in training a group of publicly released models. This enables us to collect member models in the wild and, thus, assess our method in realistic settings. After broad searching, we identify four target datasets, including databricks-dolly-15k, alpaca, SlimOrca, and OpenHermes-2.5. As summarized in ~\autoref{tab:dataset}, they have different owners and sizes. Due to the intensive amount of experiments and limited GPU budget, we randomly sample 50,000 data points for our evaluation if the dataset is larger than that.}

\revise{\noindent\textbf{Target Datasets:} To evaluate the performance of our method, we collect four victim datasets from Hugging Face, including databricks-dolly-15k, alpaca, SlimOrca, and OpenHermes-2.5. As summarized in~\autoref{tab:dataset}, they have different owners and sizes. Due to the intensive amount of experiments and limited GPU budget, we randomly sample 50,000 data points for our evaluation if the dataset is larger than that. }

\vspace{1em}
\noindent\textbf{Suspect Models:} We prioritize suspect models from the wild to better mirror the reality. Datasets released on Hugging Face comes with a card showing the list of public models fine-tuned on them (e.g.,~\cite{Databricks_dataset}). Based on that, we identify 10 member models called \textit{Hugging Face} member models. 
Similarly, we collect models released on Hugging Face but not included in a dataset's card to work as non-member models. Non-member models are more prevalent and we identify 20 of them for each target dataset. Details of the member and non-member models are summarized in~\autoref{tab:app:models}.

To increase the number of member models to balance with the non-member models, we further prepare 10 \textit{local} member models for each dataset atop widely-used models presented in~\autoref{tab:local_models}. Specifically, we fine-tune the models with QLORA~\cite{qlora} on each target dataset to create new member models. The parameters of QLORA we adjusted include \textit{rank of matrices} (8), \textit{scaling factor} (32), \textit{dropout probability} (0.05), \textit{bias type} (none), \textit{optimizer} (\texttt{paged\_adamw\_8bit}), \textit{learning rate} ($1 \times 10^{-4}$), \textit{batch size} (8), \textit{training epoch} (3)\footnote{Smaller models need more training epochs to reach acceptable training loss. Thus, for gemma-2b and glm-2b, we set their training epoch to 4.}, \textit{target modules} (all non-linear hidden layers). All other parameters are set to their default values.
Due to the limited computational resources, we use these hyperparameters instead of those suggested in~\cite{qlora}. However, these differences do not affect the results of our method. We obtain the same detection results using the suggested parameters by~\cite{qlora} as shown in \autoref{fig:suspect_para}.

\vspace{1em}
\noindent\textbf{Reference Models:} We build our reference models on top of five popular models, including Mistral-7B-v0.1~\cite{jiang2023mistral}, gemma-7b~\cite{team2024gemma}, Meta-Llama-3-8B~\cite{llama3modelcard}, Qwen2-7B~\cite{qwen2}, and glm-4-9b~\cite{glm2024chatglm}. Specifically, we take their corresponding instruct models which can better follow instructions as the non-member reference models $\mathbf{R}$. We further fine-tune the non-member models on the target dataset to produce the member reference models $\mathbf{R}^{t}$. The fine-tuning follows the same procedure as training our local member models.


\vspace{1em}
\noindent \textbf{Dataset Inference (Non-IID):} We apply our method, leveraging the above reference models, to detect non-member and member models on each dataset. We measure the recall, precision, and F1 score of the results.\footnote{Recall = \( \frac{TP}{TP + FN} \), Precision = \( \frac{TP}{TP + FP} \), F1-Score = \( 2 \times \frac{\text{Precision} \times \text{Recall}}{\text{Precision} + \text{Recall}} \), where TP, FP, and FN represent true positives, false positives, and false negatives, respectively.}

\vspace{1em}
\noindent\textbf{Dataset Inference (IID):} In the evaluation above, the non-member suspect models are trained on data irrelevant to the victim dataset. Maini et al.~\cite{maini2024llm} pointed out a more challenging situation where the training of the non-member models involves a dataset that is independent and identically distributed (IID) with respect to the victim dataset. 

\delete{We extend an evaluation to simulate the IID situation. Specifically, we evenly split a target dataset $\mathcal{D}$ into $\mathcal{D}_{x}$ and $\mathcal{D}_{y}$ and de-duplicate highly similar samples between them\footnote{We exclude a data sample from $\mathcal{D}_{y}$ if its input and output both have a 0.8+ Jaccard similarity~\cite{jaccard} with another sample in $\mathcal{D}_{x}$.}. $\mathcal{D}_{x}$ is considered as the victim dataset and adopted to prepare the reference models. The non-members presented in~\autoref{tab:app:models} are fine-tuned on $\mathcal{D}_{y}$ to work as non-member models. We further fine-tune the five reference models on $\mathcal{D}_{y}$ to obtain extra non-member models. This aims to assess the cases where the reference models and the non-members overlap and, thus, carry higher detection difficulties. In this evaluation, we only focus on detecting non-members, as the IID setting makes no difference to members.}

\revise{We extend our evaluation to simulate an IID setting. Specifically, we evenly split the original dataset $\mathcal{D}$ into two subsets, $\mathcal{D}{x}$ and $\mathcal{D}{y}$, and remove highly similar cross-subset samples\footnote{A sample in $\mathcal{D}{y}$ is excluded if both its input and output exhibit a Jaccard similarity greater than 0.8~\cite{jaccard} with any sample in $\mathcal{D}{x}$.}. We treat $\mathcal{D}{x}$ as the victim dataset, which is used to train both the member reference models and the member suspect models. The subset $\mathcal{D}{y}$, drawn from the same distribution as $\mathcal{D}_{x}$, is used to train non-member suspect models.}

\revise{In this evaluation, detecting members can be omitted because the member-detection workflow and target are identical in both IID and non-IID settings. For non-member detection, we fine-tune the Hugging Face models listed in~\autoref{tab:app:models} on $\mathcal{D}{y}$ to construct non-member suspect models. Additionally, we fine-tune the five reference models on $\mathcal{D}{y}$ to obtain extra non-member suspect models. This setup allows us to examine scenarios where the reference models and suspect models highly overlap, thereby increasing the difficulty of distinguishing them.}

\vspace{1em}
\noindent\textbf{Baseline:} To the best of our knowledge, there are no existing methods for black-box dataset inference. Yet, DPDLLM~\cite{dpdllm} \revise{and MIAVLM~\cite{hu2025membership}} offers black-box membership inference of individual data samples against LLMs and VLMs, which we adapt to provide dataset inference.

DPDLLM assumes the availability of some member samples (involved in training the suspect model) and some non-member samples (not involved in training the suspect model). Similar to our method, it involves a reference model which is fine-tuned on the member samples. Given the question of each member and non-member sample, DPDLLM obtains a response (represented as a sequence of tokens) from the suspect model and measures the probabilities for the reference model to output the same sequence of tokens. The probabilities are used as features to train a binary classifier for detecting member samples.

\revise{
MIAVLM leverages the variance in temperature sensitivity between member and non-member data for detection. Specifically, it probes the target model at varying temperatures and measure the rate of response divergence. Member sets exhibit significantly higher sensitivity, manifesting as rapid similarity degradation compared to non-members. These dynamic, set-level spectral features are utilized to train a inference classifier.}

In practical blackbox settings, the suspect model is private, preventing the acquisition of member samples and non-member samples. We bypass this restriction in our evaluation as follows. If the suspect model is a member model (i.e., trained on the victim dataset), we randomly pick 1,000 samples from the victim dataset as member samples. Otherwise, we fine-tune the suspect model with another dataset and pick 1,000 random samples from that dataset. To obtain non-member samples, we randomly select 1,000 samples from a dataset not used in training the suspect model.

To perform dataset inference, we run the binary classifier on all samples in the target dataset. If more samples are detected as member samples, we report the suspect model as a member model. Otherwise, we report it as a non-member model.

\begin{table}[htbp]
\centering
\renewcommand{\arraystretch}{1.2}
\setlength\tabcolsep{4.3pt}
\footnotesize
\caption{Dataset inference results under the non-IID setting. \texttt{HF} refers to models from Hugging Face. A value \texttt{x/y} indicates that \texttt{x}  models are correctly identified as members/non-members among \texttt{y} models.\label{tab:eval:benchmark}}
\begin{tabular}{l|l|c|c|c|c}
\toprule
 \multicolumn{2}{c|}{\textbf{Dataset}}           & \textbf{Databricks} & \textbf{Alpaca} & \textbf{Slimorca} & \textbf{Openhermes} \\
\toprule
\multirow{6}{*}{\rotatebox[origin=c]{90}{\textbf{DPDLLM}}}   & Local Members            & 4/10           & 2/10       &  6/10        & 2/10           \\
\cline{2-6}
                            & HF Members     & 10/10           & 6/10       & 7/10         &  7/10          \\
\cline{2-6}
                            & HF Non-members & 15/20           & 10/20       & 6/20         &  8/20          \\
\cline{2-6}                           
                            & Recall (\%)                   &   70.0         &  40.0      &   65.0       &    45.0       \\
\cline{2-6}                            
                            & Precision (\%)                &    73.7        &  44.4       &  48.2      &    42.9        \\
\cline{2-6}
                            & F1 (\%)             &   71.8        &  42.1      &  55.3       &   43.9         \\
\cline{1-6}                 
\multirow{6}{*}{\rotatebox[origin=c]{90}{\textbf{Our Method}}} & Local Members            &10/10            &10/10        &10/10          &10/10            \\
\cline{2-6}
                            & HF Members     &10/10            &10/10        &10/10          &10/10            \\
\cline{2-6}                           
                            & HF Non-members &20/20            &20/20        &20/20          &20/20            \\
\cline{2-6}                            
                            & Recall (\%)                & 100.0            & 100.0        & 100.0          & 100.0            \\
\cline{2-6}                          
                            & Precision (\%)                   & 100.0            & 100.0        & 100.0          & 100.0           \\
\cline{2-6}
                            & F1 (\%)                   & 100.0           &  100.0      & 100.0        &  100.0          \\
\bottomrule                     
\end{tabular}
\end{table}

\subsection{Dataset Inference Accuracy}

\begin{figure}[t]
    \centering
    
    \begin{subfigure}{\columnwidth}
        \centering
        \includegraphics[width=\columnwidth]{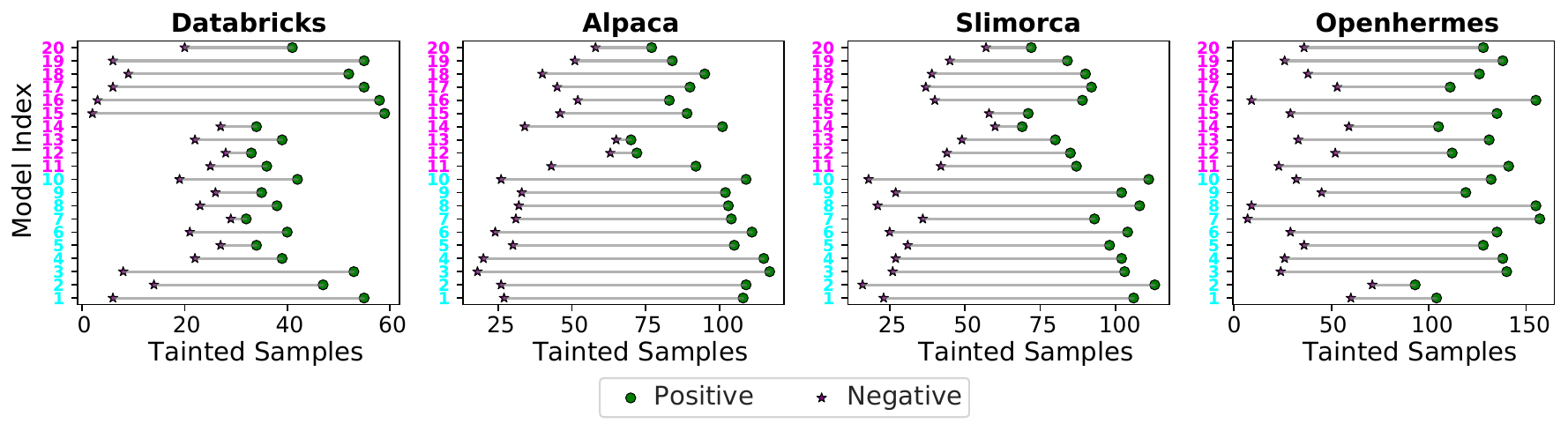}
        \caption{Distribution of positive tainted samples \textit{v.s.} negative tainted samples detected by our method for \textit{member} models. Model indexes 1-10 
        are local member models, and 11-20 are Hugging Face models.}
        \label{fig:members-noniid-dist}
    \end{subfigure}
    
    \vspace{1em} 
    
    \begin{subfigure}{\columnwidth}
        \centering
        \includegraphics[width=\columnwidth]{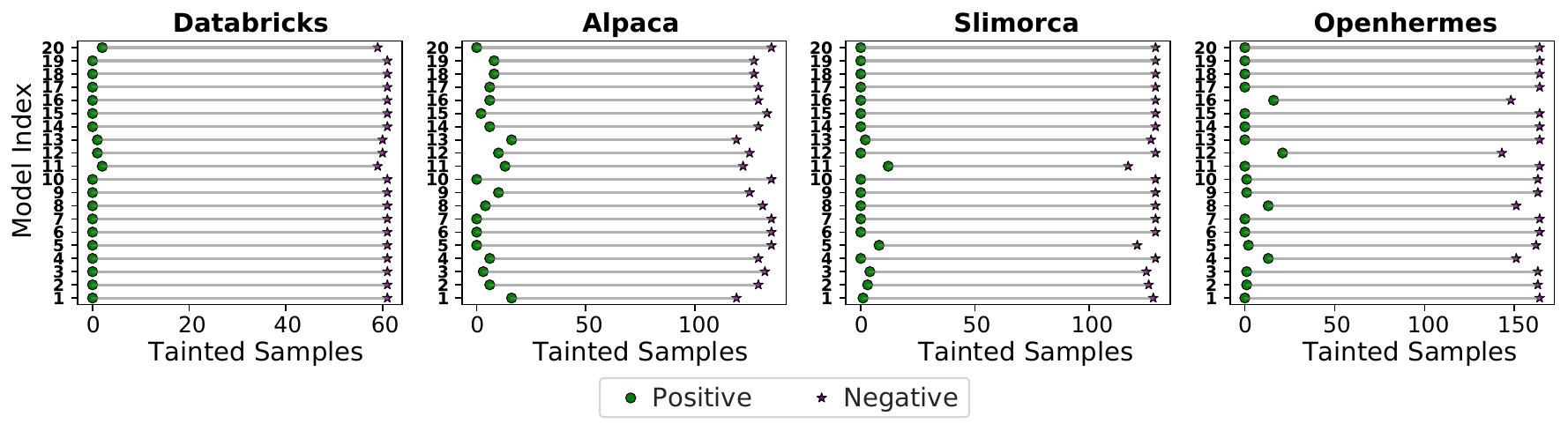}
        \caption{Distribution of positive tainted samples \textit{v.s.} negative tainted samples detected by our method for \textit{non-member} models.}
        \label{fig:non-members-noniid-dist}
    \end{subfigure}
    \caption{Comparison of positive tainted samples and negative tainted samples identified by our method for \textit{member} models (top) and \textit{non-member} models (bottom) under \textit{non-IID} settings.}
    \label{fig:ourmethod-tainted-dist}
\end{figure}

\noindent\textbf{Non-IID:} The dataset inference results under Non-IID settings are summarized in~\autoref{tab:eval:benchmark}. Our method presents 100\% accuracy, regardless of the victim dataset and the suspect model. Fundamentally, training with $\mathcal{D}$ leads the suspect model to behaving closer to the \delete{trained} \revise{member} reference models on tainted samples. As shown in the top half of~\autoref{fig:ourmethod-tainted-dist}, this enables our method to see a larger group of positive tainted samples and identify the model as a member. In contrast, the suspect model---without training on $\mathcal{D}$---diverges from the \delete{trained} \revise{member} reference models on tainted samples. Thus, our method reports most tainted sample to be negative and clearly detects the model as a non-member, as shown in the bottom half of~\autoref{fig:ourmethod-tainted-dist}.

\begin{figure}[t]
    \centering
    \begin{subfigure}{\columnwidth}
        \centering
        \includegraphics[width=\columnwidth]{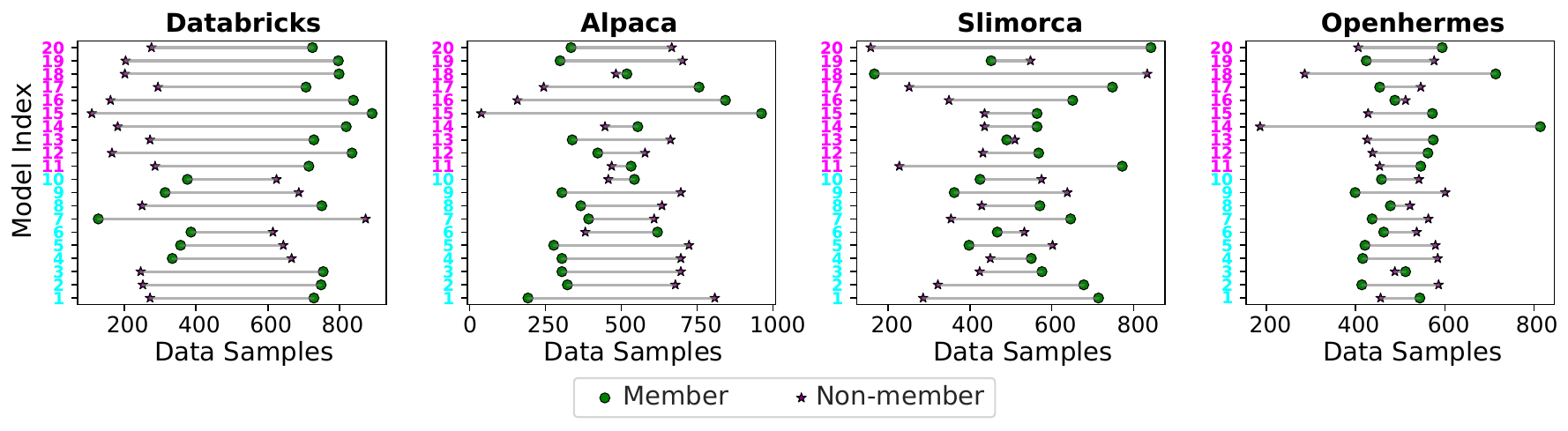}
        \caption{Distribution of member samples \textit{v.s.} non-member samples detected by DPDLLM for \textit{member} models. Model indexes 1-10 are local member models, and 11-20 are Hugging Face models.}
        \label{fig:members-noniid-dist-baseline}
    \end{subfigure}
    
    \vspace{1em} 
    
    \begin{subfigure}{\columnwidth}
        \centering
        \includegraphics[width=\columnwidth]{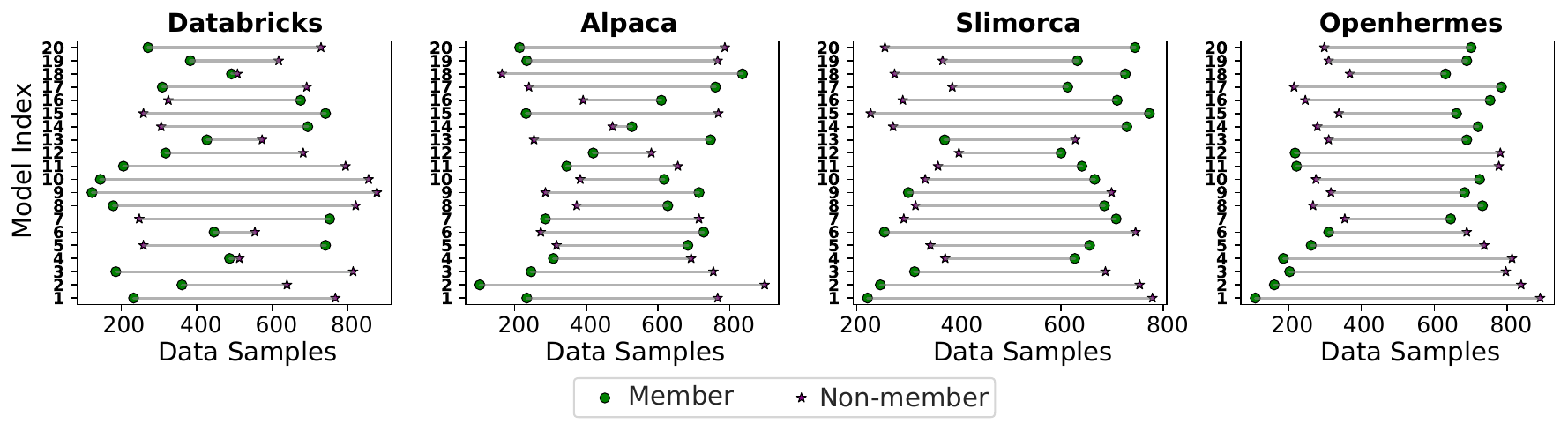}
        \caption{Distribution of member samples \textit{v.s.} non-member samples detected by DPDLLM for \textit{non-member} models.}
        \label{fig:non-members-noniid-dist-baseline}
    \end{subfigure}
    \caption{Comparison of member samples and non-member samples identified by DPDLLM for \textit{member} models (top) and \textit{non-member} models (bottom) under \textit{non-IID} settings.}
    \label{fig:baseline-tainted-dist}
\end{figure}

The baseline, DPDLLM, faces problems detecting both member and non-member models, as shown in~\autoref{tab:eval:benchmark}. On the Alpaca and Openhermes datasets, the accuracy drops to 42.1\% and 43.9\%. Even the best accuracy, achieved on Databricks, is only 71.8\%. The reason, as visualized in~\autoref{fig:baseline-tainted-dist}, is that DPDLLM presents limited accuracy when detecting the membership of individual samples. Given a member model, DPDLLM is supposed to detect all samples in $\mathcal{D}$ as members. Yet, it frequently reports more non-member samples (see top half of~\autoref{fig:baseline-tainted-dist}). Similarly, DPDLLM can mistakenly report more samples in $\mathcal{D}$ as members, given a non-member model (see bottom half of~\autoref{fig:baseline-tainted-dist}). Such observations are consistent with the findings by Maini et. al.~\cite{maini2024llm}.

\begin{table}[!t]
\centering
\renewcommand{\arraystretch}{1.36}
\footnotesize
\caption{Dataset inference results under the IID setting. ``Non-members (1)'' represent local, non-member models that share the architectures of the reference models. ``Non-members (2)'' are wild, non-member models with architectures different from the reference models. A value ``x/y'' indicates that ``x'' out of ``y'' members or non-members are correctly identified.\label{tab:eval:IID}}
\setlength\tabcolsep{1.1pt}
\begin{tabular}{l|l|c|c|c|c}
\toprule
 \multicolumn{2}{c|}{\textbf{Dataset}}           & \textbf{Databricks} & \textbf{Alpaca} & \textbf{Slimorca} & \textbf{Openhermes} \\
\toprule
\multirow{2}{*}{\textbf{DPDLLM}}   

& Non-members (1)   & 3/5           & 3/5       & 2/5         &  4/5            \\ \cline{2-6}

& Non-members (2) & 15/20           & 10/20       & 13/20         &  16/20          \\

\cline{1-6}     

\multirow{2}{*}{\textbf{Our Method}} 


& Non-members (1)    & 5/5           & 5/5       & 5/5         &  5/5            \\ \cline{2-6}

& Non-members (2) & 20/20           & 20/20       & 18/20         &  17/20          \\
\bottomrule                          
\end{tabular}

\end{table} 

\begin{figure}[!t]
    \centering
    \includegraphics[width=0.475\textwidth]{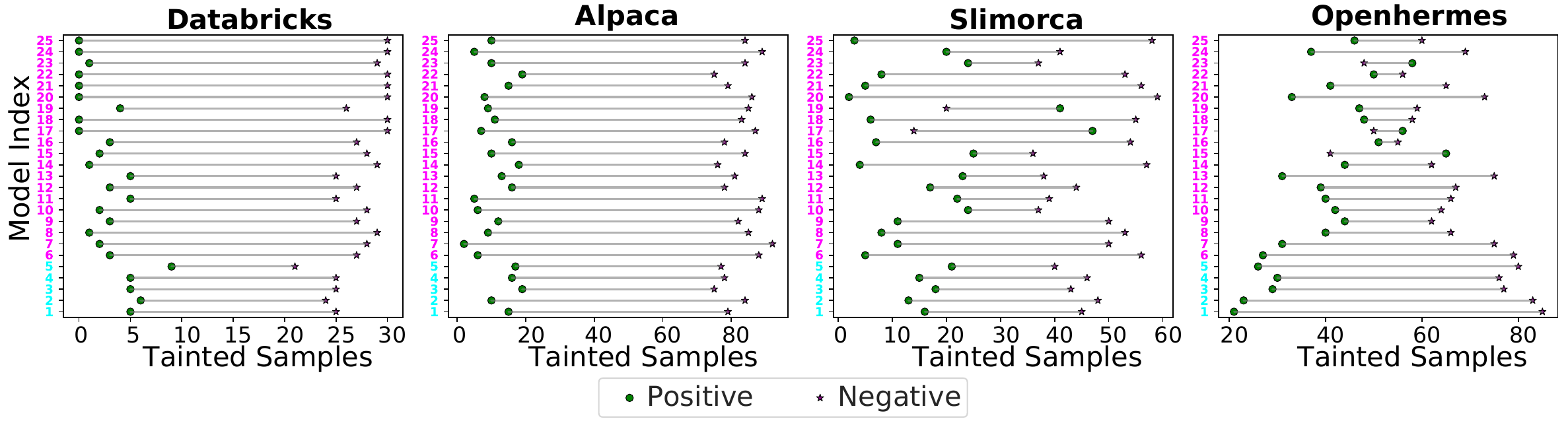}
    \caption{Distribution of positive tainted samples and negative tainted samples identified by our method for \textit{non-member} models under \textit{IID} settings.}
    \label{fig:nonmem-iid-ourmethod}
\end{figure}

\vspace{1em}
\noindent\textbf{IID:} The dataset inference results under IID settings are presented in~\autoref{tab:eval:IID}. Our method largely maintains its performance, correctly detecting all non-member models in most cases. The only exceptions occur when handling Hugging Face non-member models with Slimorca and Openhermes as the victim datasets. We mistakenly report 2 and 3 out of 20 non-members as members. As shown in~\autoref{fig:nonmem-iid-ourmethod}, our method tends to incorrectly detect more positive tainted samples, given the two datasets under the IID setting. This problem can be alleviated by including more reference models and diversifying their architectures, as we will showcase in~\S\ref{subsec:designfactors}.

\begin{figure}[!t]
    \centering
    \includegraphics[width=0.475\textwidth]{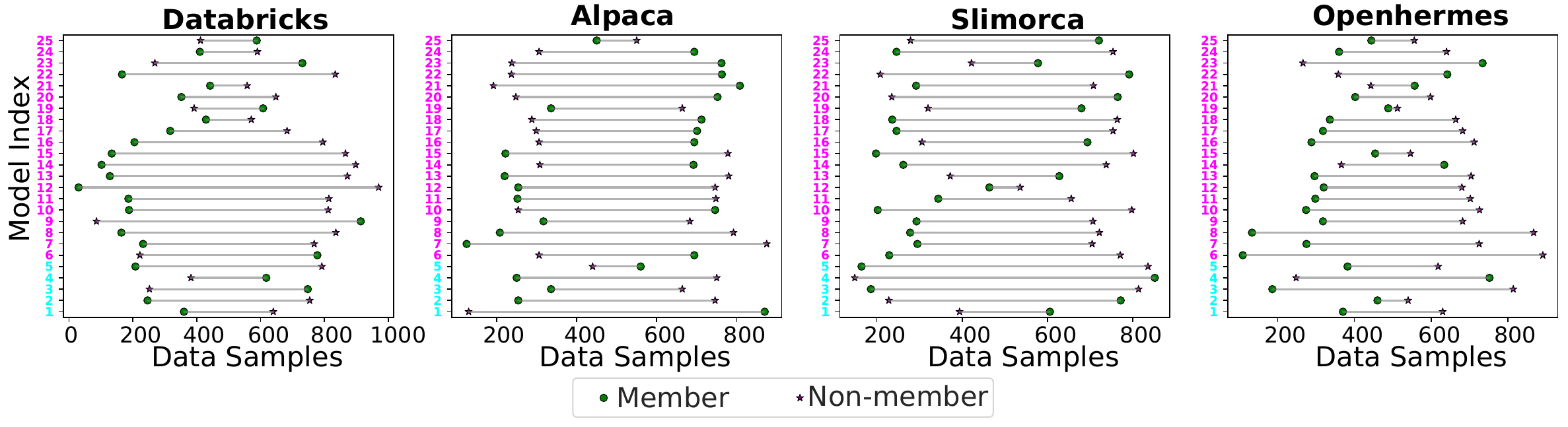}
    \caption{Distribution of member samples and non-member samples identified by DPDLLM for \textit{non-member} models under \textit{IID} settings.}
    \label{fig:nonmem-iid-baseline}
\end{figure}

In contrast, DPDLLM fails again to accurately detect non-member models under the IID setting. On the datasets of Alpaca and Slimorca, its performance approximates that of a random guess. On Databricks and Openhermes, its false positive rate is also close to 30\%. Similar to the non-IID setting, such results are attributed to DPDLLM's limited accuracy in detecting the non-member samples, as illustrated in~\autoref{fig:nonmem-iid-baseline}.

\revise{For MIAVLM, we focus on the \textit{Shadow Model Inference} attack introduced in \cite{hu2025membership}, which follows the same threat model as our work and represents the strongest approach. We observed that this approach requires a large amount of data to effectively train its MIA classifier. Specifically, it relies on response features from suspect models queried under multiple temperature values. When testing suspect models that share the same architecture as the shadow models (used to train the MIA classifier) and using responses collected at ten different temperature settings, the classifier achieved 94\% accuracy. However, when the method has access to responses obtained from only two temperatures, the accuracy drops to 67\%. In a black-box setting, the Arbiter typically does not have full access to or control over the temperature parameter of suspect models. Moreover, when the shadow models differ in architecture from the suspect models—which is particularly common in black-box scenarios—the accuracy degrades to random guess (50\%), rendering the method ineffective. Due to these limitations, we do not include it in other evaluations.}

\subsection{Impacts of Method Designs\label{subsec:designfactors}}

\subsubsection{\textbf{Tainted Samples}} 
\label{sec:tainted_samples}
A key step of our method is selecting tainted samples. Yet, its necessity needs validation. Accordingly, we re-run the evaluations under the non-IID settings with our method applied to all samples in the target dataset.

As shown in~\autoref{tab:eval:all_samples}, our method fails when all samples are used. Specifically, our method will detect every model as a non-member, regardless of the ground truth. The results are not surprising. Most samples in $\mathcal{D}$ do not impact the behaviors of the reference models. Hence, the similarity between the suspect model and the reference models on those samples doesn't vary much before and after we fine-tune the reference models. As a demonstration, ~\autoref{fig:all-sample-hist} shows the distribution of this similarity difference across all samples, using a member model from Hugging Face as the suspect model. Yet, our method requires a sample to stay closer to the member reference models by a margin of $\delta^s$ to be considered a positive tainted sample. As a result, our method end up detecting most samples as negative tainted samples, consistently reporting the suspect model to be a non-member.

\revise{Additionally, we evaluate how tainted samples affect the performance of the baseline method DPDLLM. The results in \autoref{tab:baseline_tainted} show that using tainted samples in DPDLLM does not improve its accuracy. This is because DPDLLM needs to train a binary classifier for detection. Using only tainted samples provides too little data to train a well-performing classifier, and therefore it cannot achieve good detection accuracy.}

\begin{figure}[htbp]
    \centering
    \includegraphics[width=0.475\textwidth]{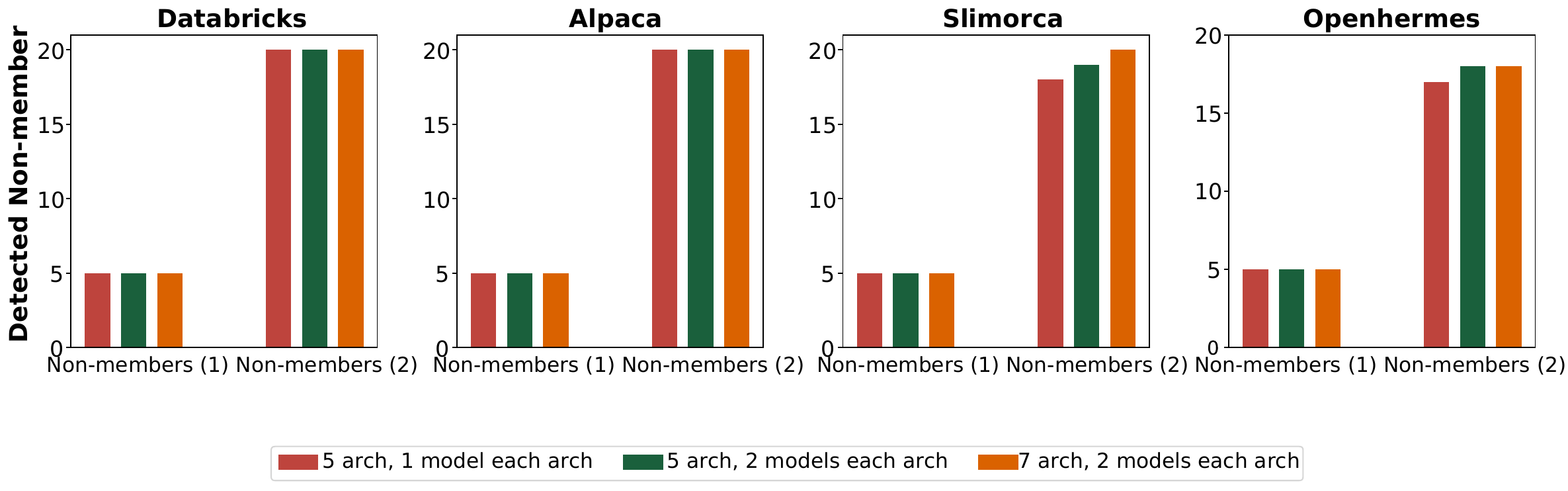}
    \caption{Impacts of reference model number and architecture diversity on our method under IID settings. ``Non-members (1)'' represent local, non-member models that share the architectures of the reference models. ``Non-members (2)'' are wild, non-member models with architectures different from the reference models. }
    \label{fig:ref-numer-arch}
\end{figure}

\subsubsection{\textbf{Reference Models}}
\label{sec:reference_models}
Our method relies on reference models to work. By intuition, the number and diversity of reference models can make a difference. We extend evaluations to understand their impacts as follows.

\vspace{0.5em}
\textbf{Quantity:} All our evaluations so far only used one reference model each architecture. While achieving perfect accuracy under the non-IID setting, it mis-detects several non-member models under the IID setting. In this evaluation, we explore whether using more reference models improve our accuracy under the IID settings. Specifically, we increase the number of reference model to two per architecture  --- one fine-tuned with the parameters described in~\S\ref{sec:setup} and another using \textit{rank of matrices} (4), \textit{scaling factor} (16), \textit{dropout probability} (0.05), \textit{bias type} (none), \textit{optimizer} (\texttt{paged\_adamw\_8bit}), \textit{learning rate} ($5 \times 10^{-5}$), \textit{batch size} (8), \textit{training epoch} (4), \textit{target modules} (all non-linear hidden layers).

We present the results in~\autoref{fig:ref-numer-arch}. Using more reference models enabled us to detect one extra non-member model on the Slimorca dataset and another extra on the Openhermes dataset, without introducing side effects. To reason why the improvement happens to those models, we visualize the impact of the reference model number on the distribution of tainted samples in~\autoref{fig:ref-tainted-change}. Evidently, the additional reference models dramatically decreased the number of positive tainted samples, which eventually corrected our inference decisions. Such results are well expected. When more reference models are in effect, both~\autoref{equation1} and ~\autoref{equation2} become harder to satisfy, more effectively filtering out ``fake'' positive tainted samples.\looseness=-1

\vspace{0.5em}
\textbf{Diversity:} Going beyond increasing the model number, we further add two architectures, Qwen1.5-7B and glm-10b, for the reference models. To see the accumulative effects of more numbers and more architectures, we prepare two reference models for each architecture, following the configurations we described above.

As shown in~\autoref{fig:ref-numer-arch}, diversifying the architectures enabled our method to additionally detect a non-member model (Platypus2-13B) on the Slimorca dataset. As illustrated in~\autoref{fig:ref-tainted-change}, this is also attributed to that the extra reference models make both~\autoref{equation1} and ~\autoref{equation2} harder, which eventually rules out  “fake” positive
tainted samples.

\vspace{0.5em}
\textbf{Hyperparameters:} We also explore the impact of hyperparameters used to train the reference models. Our default hyperparameters are \textit{batch size} = 8 and \textit{training epoch} = 3. We include four more groups of settings with increasing \textit{batch sizes} (16, 32) and decreasing \textit{training epochs} (1, 2), respectively.

\begin{figure}[t]
    \centering
    \begin{subfigure}{\columnwidth}
        \centering
        \includegraphics[width=\columnwidth]{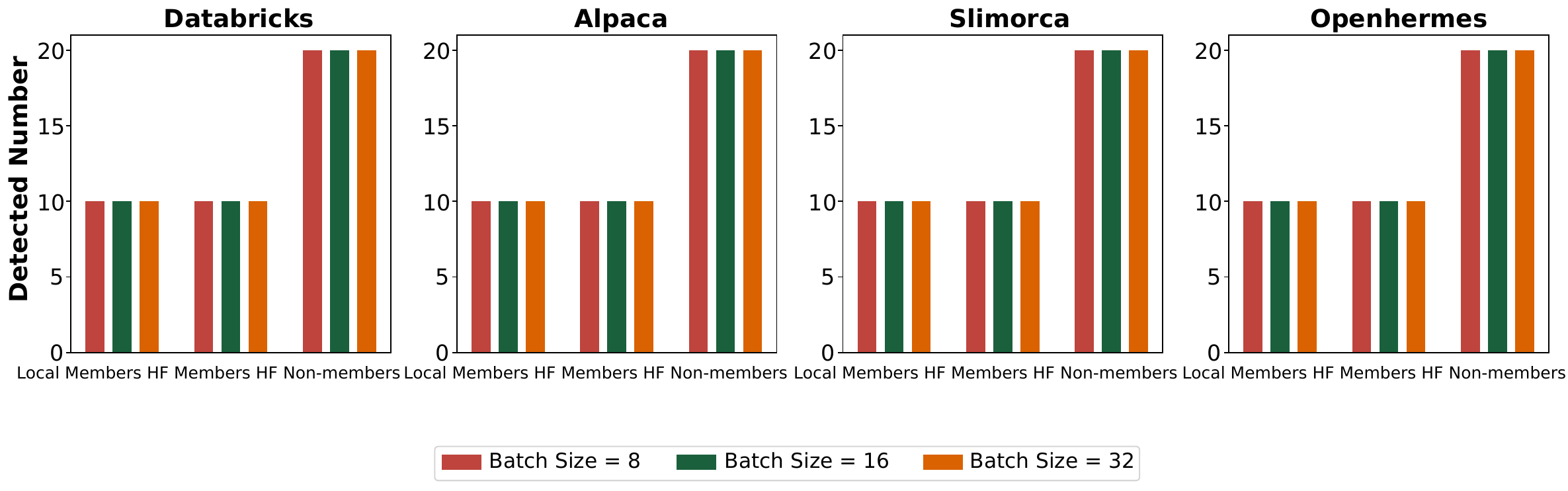}
        \caption{Impact of batch size under Non-IID settings.}
        \label{fig:ref_batch_nonIID}
        \vspace{0.5em}
    \end{subfigure}
    \begin{subfigure}{\columnwidth}
        \centering
        \includegraphics[width=\columnwidth]{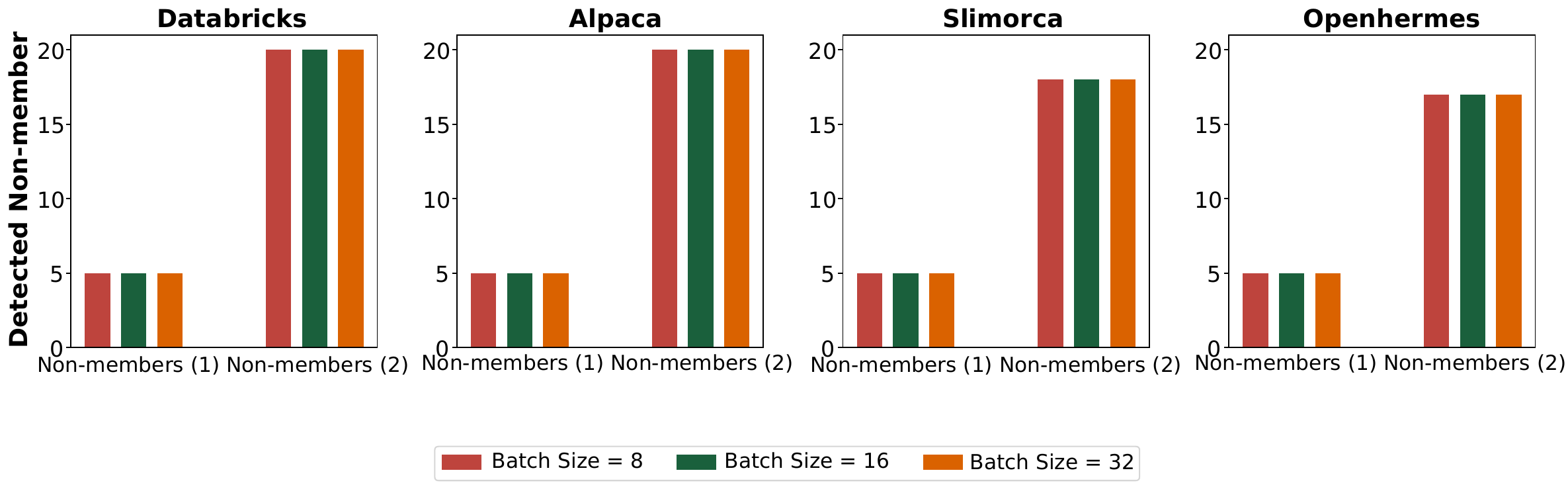}
        \caption{Impact of batch size under IID settings.}
        \label{fig:ref_batch_IID}
        \vspace{0.5em}
    \end{subfigure}
    \begin{subfigure}{\columnwidth}
        \centering
        \includegraphics[width=\columnwidth]{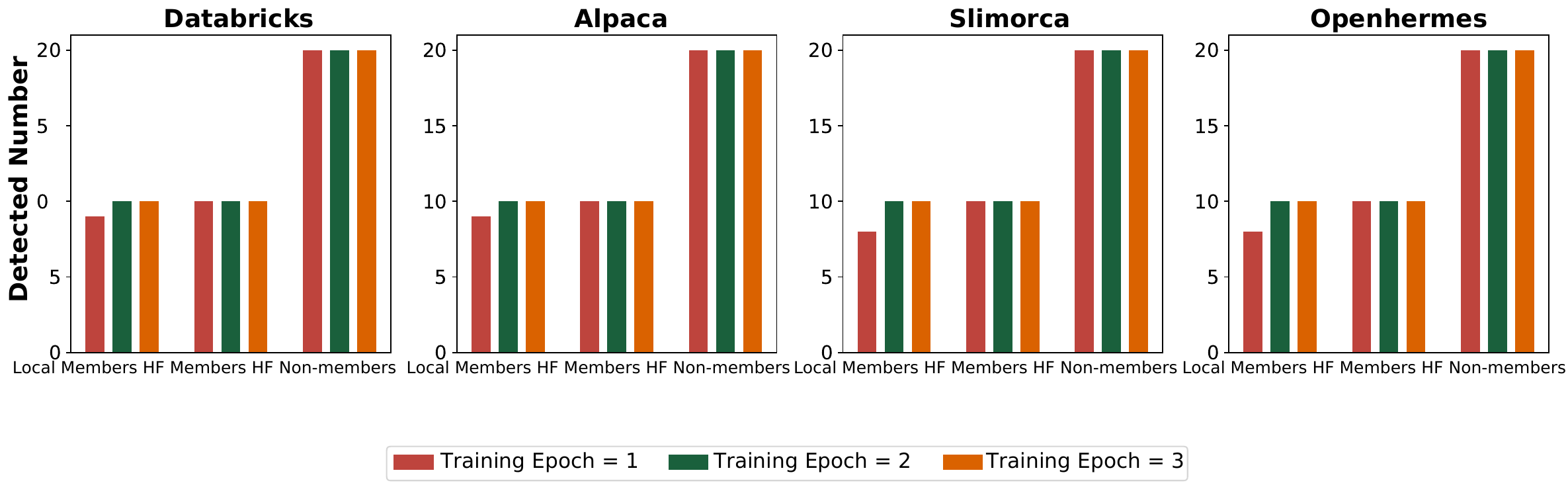}
        \caption{Impact of training epoch under Non-IID settings.}
        \label{fig:ref_epoch_nonIID}
        \vspace{0.5em}
    \end{subfigure}
    \begin{subfigure}{\columnwidth}
        \centering
        \includegraphics[width=\columnwidth]{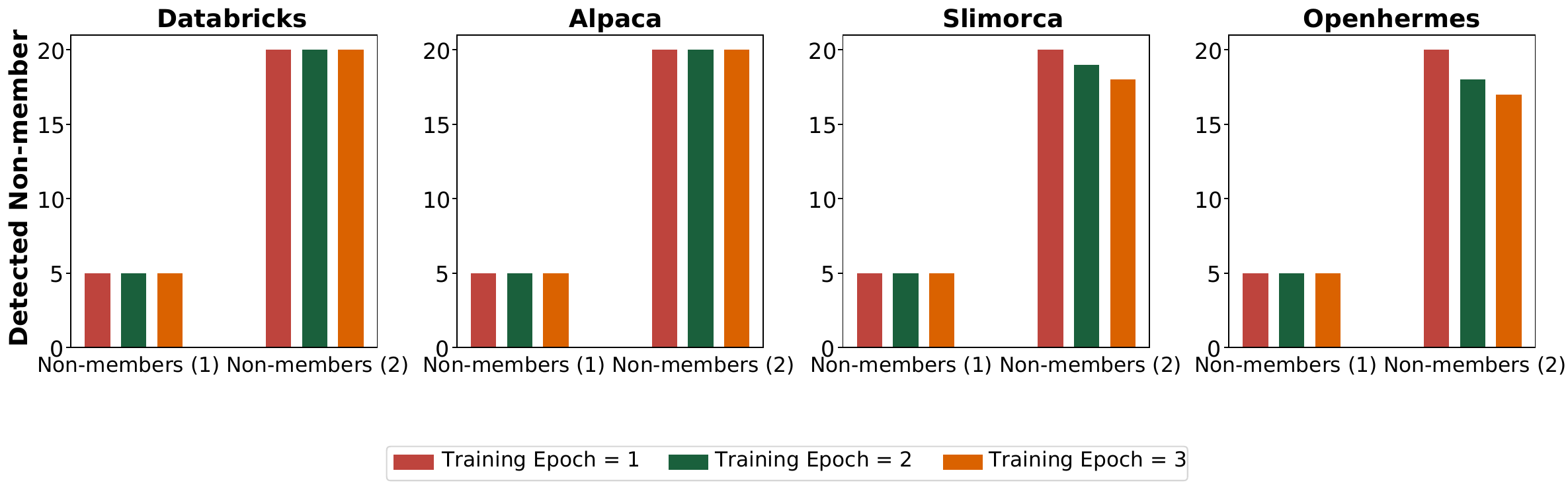}
        \caption{Impact of training epoch under IID settings.}
        \label{fig:ref_epoch_IID}
        \vspace{0.5em}
    \end{subfigure}
    \caption{Impact of hypeparameters for reference models on dataset inference. ``HF'' refers to Hugging Face. ``Non-members (1)'' represent local, non-member models that share the same architectures of the reference models. ``Non-members (2)'' are wild, non-member models with architectures different from the reference models.}
    \label{fig:ref_para}
\end{figure}

As shown in \autoref{fig:ref_para}, changing the \textit{batch size} does not affect the accuracy of dataset inference, as it does not influence the learning effectiveness of the reference models.
However, a small number of training epochs leads to slightly lower detection rates of suspect models due to insufficient learning.
Under the non-IID setting, a few local members are incorrectly detected, though this does not affect detection under the IID setting.
More training epochs, on the other hand, can lead to the incorrect identification of certain non-members.
To understand the underlying cause, we visualize the impact of training epochs of reference models on the distribution of tainted samples in \autoref{fig:ref_epoch_tainted}.
We observe that increasing the number of training epochs enhances the reference models' ability to select more true positive tainted samples and filter out more false positives under non-IID settings for local members.
However, it has the opposite effect under the IID setting for non-members.

\vspace{0.5em}
\textbf{Discussion:} Our evaluations above suggest that using more reference models with diversified architectures enhances the accuracy of our method. Yet, it does not mean we should endlessly increase reference models and their architectures. A clear trend unveiled in~\autoref{fig:ref-tainted-change} is that a larger family of reference models result in fewer tainted samples. When using 7 architectures with 2 models for each architecture, we often end up with fewer than 50 tainted samples in total. Decisions based on such a limited sample size are vulnerable to even minor disturbances. For instance, flipping of a small number of negative sample to positive ones due to randomness could overturn the inference result.
Additionally, regarding hyperparameters, the batch size has no impact on our method. However, an insufficient number of training epochs can degrade the quality of the tainted samples. Increasing the number of training epochs helps filter out false positives of tainted samples and select more true positives, and hence improves the performance of our method.

\subsubsection{\textbf{Dataset Size}} We study how the size of the dataset available for inference affects our technique. Specifically, we randomly select 5000, 10000, and 15000 of the original dataset to perform dataset inference. In order to obtain wild member models, for each dataset, we fine-tune Hugging Face non-members listed in \autoref{tab:app:models} on these subsets as Hugging Face members. Additionally, we fine-tune reference models with these subsets to select tainted samples.\label{sec:size}

As shown in \autoref{fig:dataset_trend}, non-member models are consistently identified correctly, whereas member models are slightly affected by the smaller size of the dataset. However, the impact is limited as our method can still achieve $>$95\% accuracy in most cases. With the increase of the dataset size, all member models are correctly classified. These results demonstrate the robustness of our method with respect to the size of the available dataset for inference.

\begin{figure}[t]
    \centering
    \begin{subfigure}{\columnwidth}
        \centering
        \includegraphics[width=\columnwidth]{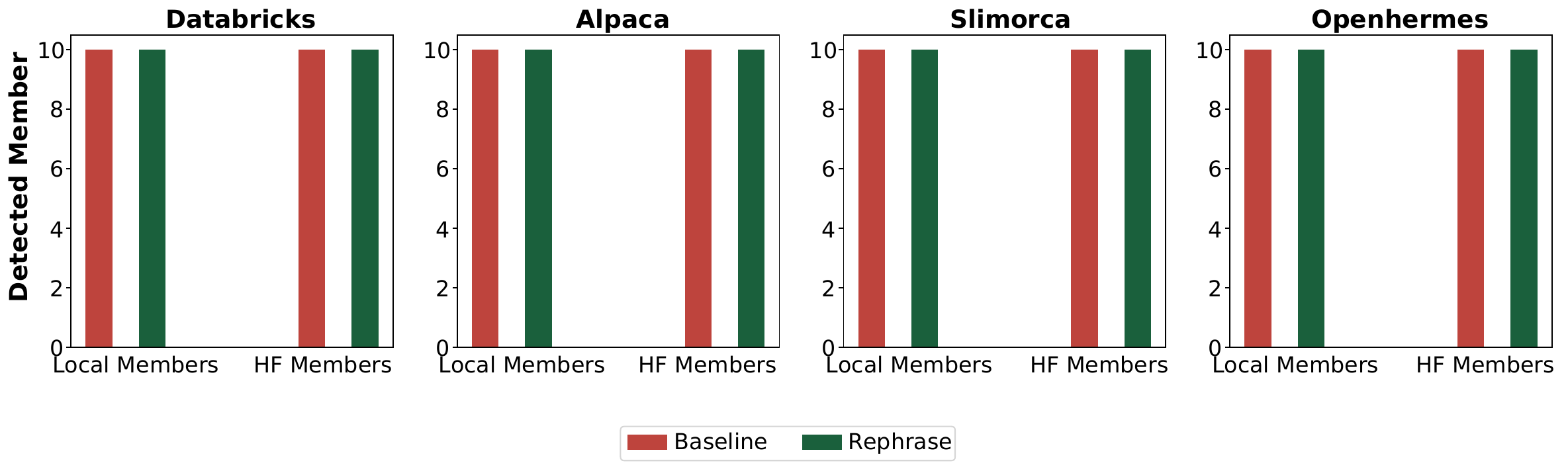}
        \caption{Against rephrasing responses.}
        \label{fig:rephrasing}
        \vspace{0.5em}
    \end{subfigure}
    \begin{subfigure}{\columnwidth}
        \centering
        \includegraphics[width=\columnwidth]{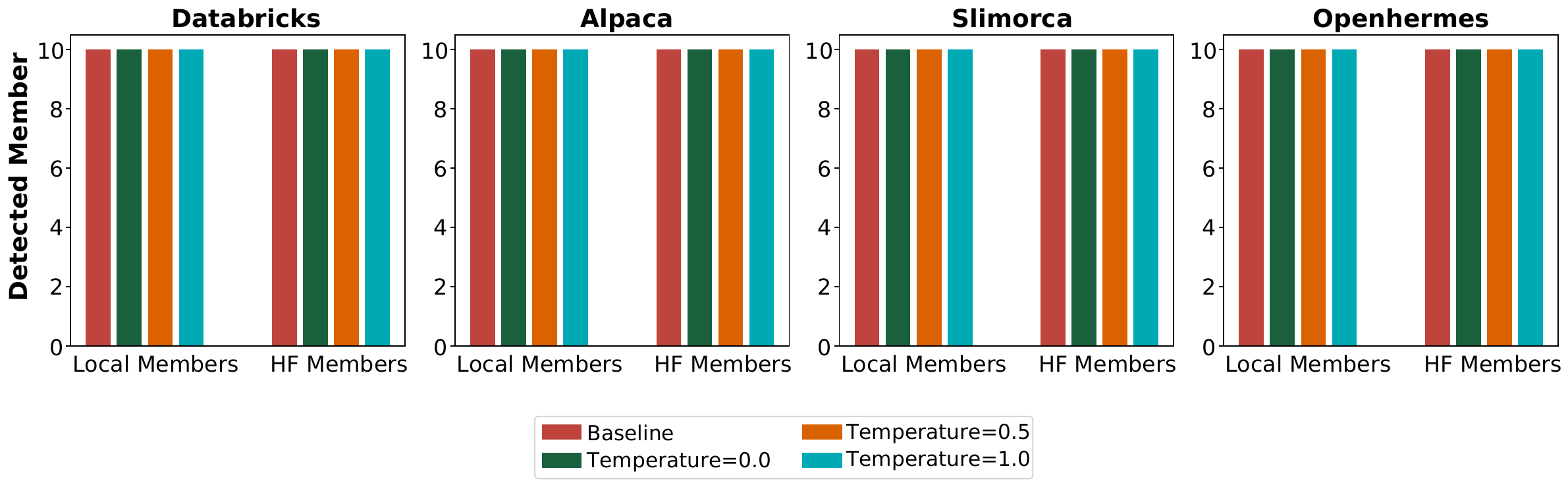}
        \caption{Against changing temperatures.}
        \label{fig:temperature}
        \vspace{0.5em}
    \end{subfigure}
    \begin{subfigure}{\columnwidth}
        \centering
        \includegraphics[width=\columnwidth]{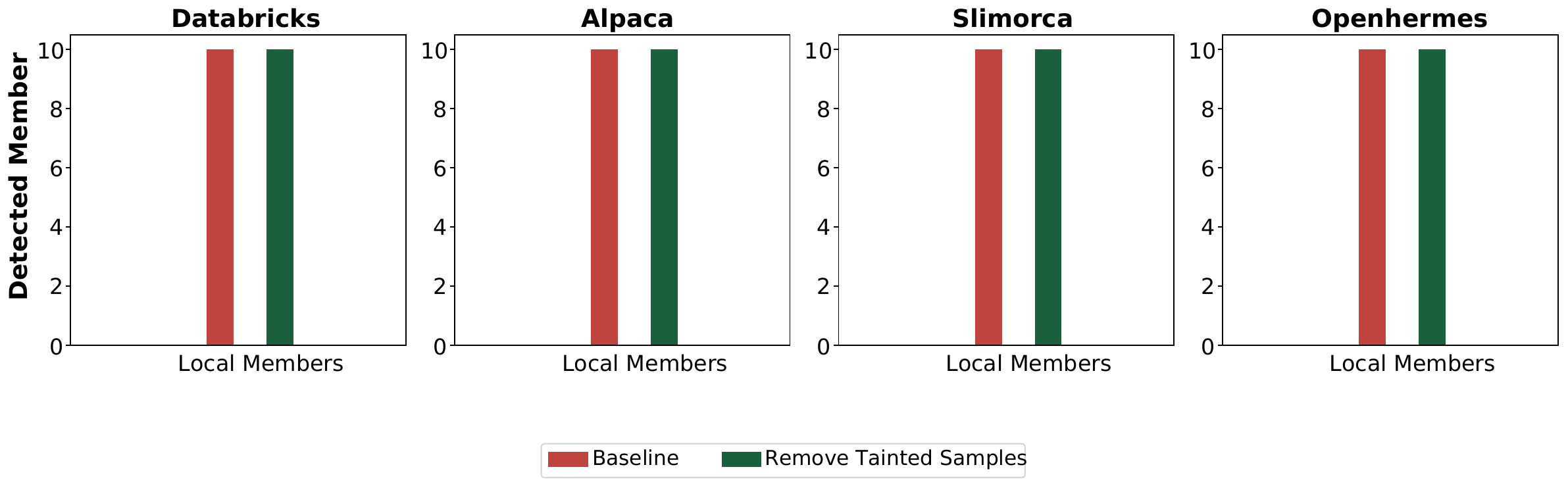}
        \caption{Against removing tainted samples.}
        \vspace{0.5em}
        \label{fig:removing}
    \end{subfigure}
    \begin{subfigure}{\columnwidth}
        \centering
        \includegraphics[width=\columnwidth]{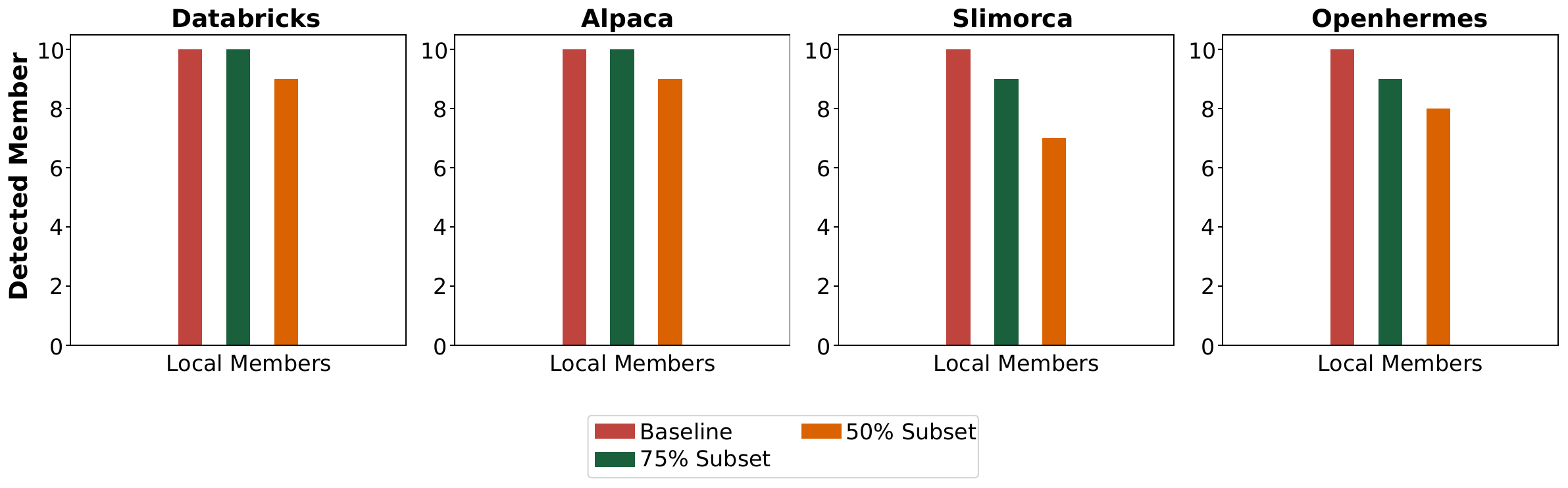}
        \caption{Against training on a subset of samples.}
        \vspace{0.5em}
        \label{fig:subset}
    \end{subfigure}
    \caption{Robustness of our methods against different adversarial evasion attempts. ``HF'' refers to Hugging Face.}
    \label{fig:adv-robustness}
\end{figure}

\subsection{Adversarial Robustness\label{subsec:robustness}}
In practice, the adversary may adopt various countermeasure to evade our dataset inference method. In this section, we evaluate the robustness of our method against several possibilities, including \textbf{rephrasing responses}, \textbf{changing temperature}, \textbf{removing tainted samples}, \textbf{training on subsets}, and \textbf{training on multiple datasets sequentially}. We focus on the member models as evasion intends to avoid the detection of them. Further, we reuse the setup discussed in~\S\ref{sec:setup} for the reference models (5 architectures, one model each architecture).

\subsubsection{\textbf{Rephrasing Responses}} Before sending responses out, the adversary can rephrase them, attempting to thwart our similarity measurement. We simulate the scenario by adopting GPT-4o for rephrasing and re-run the evaluations.

As shown in~\autoref{fig:rephrasing}, our method still detects all member models when responses are rephrased, indicating our robustness against this countermeasure. The results are attributed to our use of BERTScore, which intends to measure the semantic similarities and, thus, preserves utilities on rephrased texts.

\subsubsection{\textbf{Changing Temperature}}
\label{sec:temperature}
Temperature is a hyperparameter that influences an LLM's output by scaling the model's predicted probability distribution. A higher temperature will result in lower probability, producing more creative outputs~\cite{llmtem}. Thus, varying the temperature is expected to alter the model behaviors, which offers another evasion option to the adversaries. We experiment with this option by setting three different temperatures, including 0.0 (greedy sampling), 0.5 (half-creative sampling), and 1.0 (full-creative sampling), for the suspect model and re-perform the evaluations. All the reference models use their default temperature 1.0. 

The evaluation results are illustrated in~\autoref{fig:temperature}. Our method detects all the member models, given different temperatures. It demonstrates the robustness of our method against this evasion attempt. Presumably, two designs of our method contribute to this robustness. First, our use of BERTScore enables us to inspect sematic-level similarities, reducing the impacts of creativity on wording and phrasing. Second, we obtain multiple responses from the suspect model but only use the one leading to the largest similarity difference (see~\S\ref{subsec:des:datasetinf}). This helps reduce variations introduced by the temperature.

\subsubsection{\textbf{Removing Tainted Samples}} Adversaries can also attempt to remove the tainted samples from the victim dataset before training their models. A key issue is whether the adversaries will be granted access to the reference models. We argue that they should not because there is no necessity for the \textit{Arbiter}--who performs dataset inference--to release the reference models to third parties. If reproduction of the inference results is mandated, the \textit{Arbiter} can release the reference models after membership is detected.

In accordance to the above, our evaluation assumes that the adversaries do not have the original reference models. Instead, they create their own ones. To simulate adversaries with better luck, we assume they may use architectures similar to the real reference models. Eventually, we include Mistral-7B-v0.3, gemma-2b, Llama-3.1-8B, Qwen1.5-7B, and glm-10b as the adversarial reference models and used them to remove positive tainted samples before training the member models. By removing positive tainted samples, the \textit{Arbiter} is supposed to detect more negative tainted samples and fewer positive tainted samples, making it less possible to detect member models. Since the evaluation requires re-training, we cannot apply it to the Hugging Face models. Thus, we focus on the local member models.

~\autoref{fig:removing} presents the evaluation results, demonstrating that removing tainted samples does not diminish our detection capability. This outcome highlights the robustness of our method against such evasion attempts. The effectiveness can be attributed to the discrepancy between the reference models used by the \textit{Arbiter} and those used by the adversary, which results in minimal overlap among the identified tainted samples. \revise{There exist many tainted samples that can be used to detect the suspect model. Using different reference models to select tainted samples leads to only a very small overlap between the resulting tainted sets. Although each set of tainted samples can independently validate the suspect model, they do not share the same tainted data. Therefore, most tainted samples selected by the arbiter remain effective for detecting the suspect model and are not affected by the adversary.}

Further details on the overlap of tainted samples are illustrated in \autoref{fig:tainted-overlap}. It reveals that the evasion strategy of removing tainted samples exhibits a dual effect. On one hand, it has a seemingly beneficial impact by eliminating some positive tainted samples — for example, in the Alpaca dataset, up to 29 overlapping positive tainted samples were removed. On the other hand, it inadvertently removes negative tainted samples as well, with up to 7 such overlaps also observed in the Alpaca dataset. These opposing effects partially cancel each other out. Moreover, since the original number of positive samples significantly exceeds the number of negative samples, the resulting ratio after removal remains insufficient to alter the decision-making outcome of our method. Therefore, removing tainted samples based on the adversary's reference models does not substantially reduce the chance that suspect models will still encounter tainted samples as recognized by the \textit{Arbiter} during fine-tuning. In conclusion, concealing the reference models employed by the \textit{Arbiter} offers an effective defense against evasion via tainted sample removal.

\subsubsection{\textbf{Training On Subsets}} Instead of using the complete victim dataset, adversaries may only use a subset for training. This can create problems for our detection as we still identify tainted samples from the entire dataset. Yet, we suspect that subset detection is no longer suitable to be solved through dataset inference. In particular, we raise the question that whether subset detection and dataset inference always share the same answer. Take the extreme case where only one sample is used for training as an example. Conceptually, it is still a subset, but we can hardly say the dataset is used. In our opinion, inference of individual samples (MIA) is a more desired solution for subset detection.

Nevertheless, we extend evaluations to understand the impact of subset training on our method. We re-train the local member models using 75\% and 50\% of samples randomly picked from the victim dataset, and then re-run our dataset inference. The results are presented in~\autoref{fig:subset}. When 75\% samples are still used in training, our method largely maintains its detection capability, missing only one member on the Slimorca dataset and one on the Openhermes dataset. When the ratio of training samples drops to 50\%, our method start missing member models on all datasets. On Slimorca and Openhermes, we miss 3 out of 10. This is expected as we still consider all the data to identify tainted samples. Among those, more positive tainted samples will be excluded when a smaller subset is used for training, which turn into negative ones and eventually flip our inference result.

\subsubsection{\textbf{Training On Multiple Datasets Sequentially}} Instead of training LLMs on multiple datasets simultaneously, adversaries may fine-tune the models sequentially on one dataset at a time.
This approach might reduce the memorization effect on earlier datasets and potentially allow adversaries to bypass our inference method.
To evaluate this scenario, we train the local models in \autoref{tab:local_models} on the datasets listed in \autoref{tab:dataset} in a sequential manner.
Reference models are trained separately on each dataset to select the corresponding tainted samples.
Each time a new dataset is used for training, we perform dataset inference on all previously used datasets to determine which ones can still be successfully detected.

As shown in \autoref{tab:eval:seq_train}, our approach is effective at detecting the two most recently used training datasets. However, its effectiveness decreases for the third-oldest and the oldest datasets. To understand the cause of this decline, we evaluate the utility performance of two representative models on all the datasets in a sequential manner, as shown in \autoref{fig:utility_trend}. We observe that the models’ utility on earlier training datasets drops significantly. For example, the BERTScore of glm-2b on the oldest dataset, Databricks, decreases from nearly 1.00 to below 0.70, denoting a degradation of over 30\%. This suggests that the model may have forgotten knowledge from the dataset, making it undetectable by our method. In practice, the attack aims to fully utilize the dataset to achieve high utility; however, the above approach contradicts this goal. Therefore, as long as the member model achieves good utility performance, it can be detected by our technique.

\subsection{Extension to Additional Task}
We extend our method to the domain of text generation. For this evaluation, we select the popular \textit{C4} dataset\footnote{https://huggingface.co/datasets/allenai/c4} provided by \textit{AllenAI} on Hugging Face.
Specifically, we use the English subset and randomly select 50,000 data points as $\mathcal{D}$.
We then choose five wild member models and five wild non-member models, as described in \autoref{tab:text_gen_models}.
Following our proposed dataset inference method, we use the same non-member reference models from~\S\ref{sec:design} to construct member reference models and identify tainted samples.
The evaluation results are presented in \autoref{tab:eval:text_gen}. These results demonstrate that our method generalizes well to other tasks.

\subsection{Efficiency}

\delete{To understand the efficiency---which is important in practical use---of our method, we measure its time cost for dataset inference. Our method involves two phases: (i) an \textit{offline} phase which builds the reference models and identifies the tainted samples and (ii) an \textit{online} phase which tests a given suspect model. We separately measure the time cost of each phase on machines with an AMD EPYC 7513 32-Core Processor, an NVIDIA A100 (80 GB GPU memory), and 256 GB RAM. \autoref{tab:efficiency} shows the evaluation results. The offline phase often takes hours to accomplish due to the time-consuming process of training reference models. Yet, it only needs to run once for every dataset, incurring a limited impact in practice. More importantly, the online phase only needs around 25 minutes to finish, presenting a high efficiency in the more frequent operations.}

\revise{To evaluate the practical efficiency of our method, we measure its dataset-inference time cost across two phases: (i) an \textit{offline} phase that builds reference models and identifies tainted samples, and (ii) an \textit{online} phase that tests a suspect model. We report each phase’s runtime on a machine equipped with an AMD EPYC 7513 32-Core CPU, an NVIDIA A100 (80 GB), and 256 GB RAM. As shown in \autoref{tab:efficiency}, the offline phase takes hours due to reference-model training but is performed only once per dataset. In contrast, the online phase completes in about 25 minutes, offering high efficiency for routine use.}
\section{Related Works}

\noindent\textbf{Membership Inference Attack on LLMs:}
Membership inference (MI) identifies whether a data point was used in training, with uses in detecting benchmark contamination~\cite{oren2023proving}, auditing privacy~\cite{steinke2023privacy}, and finding copyrighted content~\cite{carlini2021extracting}. While well-studied in small models, MI for LLMs is emerging. Methods are mainly score-based or reference model-based.

\vspace{1em}
\noindent\textbf{Score-Based Membership Inference Attacks (MIAs)} assume models behave differently on training versus unseen data, which can be captured by some statistical metrics such as the perplexity~\cite{carlini2021extracting} or the loss values~\cite{yeom2018privacy}. After gathering those metrics, they compare the model's output score to the predefined threshold or use statistical methods to infer membership. Mattern et al. ~\cite{mattern2023membership} propose the neighborhood attack, which generates highly similar "neighbor" sentences using a pretrained masked language model and compares their losses to the original sample under the target model. If the original sample was part of the training data, the loss difference between it and its neighbors will be smaller than a predefined threshold, indicating membership. Min-k\% Prob~\cite{shi2023detecting} calculate the average log likelihood of the 
k\% tokens with the lowest probabilities. A high average log-likelihood suggests the text is likely part of the pretraining data, as seen examples tend to lack outlier words with very low token probabilities. However, their effectiveness depends on the assumption that the model behaves differently for training versus non-training data, which may not hold if the model generalizes well. 

\vspace{1em}
\noindent\textbf{Reference Model-Based MIAs}, on the other hand, involve training a separate reference model on a dataset similar to the target model's training data. By comparing the behavior (e.g., perplexity or loss values) of the target model and the reference model, attackers can infer membership more robustly. Carlini et al.~\cite{carlini2021extracting} compares the perplexity ratio between a suspect model (e.g., a large GPT-2 model) and a smaller reference model (e.g., a smaller model) for a given text. If the suspect model has memorized the text during training, its perplexity will be significantly lower than the reference model's, as the smaller reference model is not able to memorize as much training data as the suspect model.

\vspace{1em}
\noindent\textbf{Dataset Inference on LLM}. Dataset inference determines whether a specific dataset was used in training. As far as we know, Maini et al.~\cite{maini2024llm} present the only known work tackling this problem for LLMs. They show that different MIAs vary in performance across datasets, and no single attack performs consistently. Their approach extracts MIA-derived features from both suspect and validation datasets, trains a linear model to learn correlations, and applies a statistical T-Test to infer dataset inclusion.
\section{Conclusion}

In conclusion, this paper introduces a new method for dataset inference in a black-box setting. The method utilizes reference models to identify tainted samples and detect suspect models by comparing response similarities. Our evaluations with both wild models and self-trained models demonstrate the effectiveness of the approach. Additionally, we experimented with a series of evasion attempts against our method, showing that our method offers robustness against those attempts.


\section{Acknowledgment}

This work was supported by National Science Foundation (NSF) awards OAC-2319880, CNS-2029038 and CNS-2135988. We are also grateful to NVIDIA for providing computational resources. Any opinions, findings, and conclusions or recommendations expressed herein are those of the authors and do not necessarily reflect the views of the US government or NSF.

\newpage
\bibliographystyle{IEEEtran}
\bibliography{paper}

\clearpage

\appendix
\subsection{Results}
%
\begin{strip}
\centering
\captionsetup{type=table}
\captionof{table}{The list of member models and non-member models we collected from the Hugging Face in our evaluations.}
\vspace{-2em}
\label{tab:app:models}
\end{strip}
\begin{center}
\scriptsize
\resizebox{\textwidth}{!}{
\begin{tabular}{c|c|cc|cc}
\cline{1-6}
\multicolumn{2}{l|}{\multirow{2}{*}{\textbf{Dataset}}} & \multicolumn{2}{c|}{\textbf{Databricks}} & \multicolumn{2}{c}{\textbf{Alpaca}} \\
\cline{3-6}
\multicolumn{2}{l|}{} & Owner & Name&  Owner & Name \\
\cline{1-6}
\multirow{10}{*}{\rotatebox[origin=c]{90}{Member Models}}
& 1&  ikala&  redpajama-3b-chat&  declare-lab&  flan-alpaca-xl \\
& 2&  databricks&  dolly-v2-3b&  declare-lab&  flan-alpaca-xxl \\
& 3&  databricks&  dolly-v2-7b&  PKU-Alignment&  alpaca-7b-reproduced \\
& 4&  databricks&  dolly-v2-12b&  PKU-Alignment&  alpaca-8b-reproduced-llama-3 \\
& 5&  TheBloke&  tulu-7B-fp16&  PKU-Alignment& alpaca-7b-reproduced-llama-2 \\
& 6&  TheBloke&  tulu-13B-fp16&  GeorgiaTechResearchInstitute& galpaca-6.7b \\
& 7&  allenai&  open-instruct-pythia-6.9b-tulu&  GeorgiaTechResearchInstitute& galpaca-30b \\
& 8&  allenai&  allenai/open-instruct-human-mix-65b&  luckychao& TinyAlpaca-1.1B \\
& 9&  allenai&  open-instruct-opt-6.7b-tulu&  hiyouga& Llama-2-Chinese-13b-chat \\
& 10&  HuggingFaceH4&  starchat-alpha&  NEU-HAI& Llama-2-7b-alpaca-cleaned \\
\cline{1-6}
\multirow{20}{*}{\rotatebox[origin=c]{90}{Non-Member Models}}
& 1&  Open-Orca&  Mistral-7B-OpenOrca&  ikala&  bloom-zh-3b-chat \\
& 2&  h2oai&  h2o-danube-1.8b-chat&  databricks&  dolly-v2-3b \\
& 3&  jondurbin&  bagel-8b-v1.0&  databricks&  dolly-v2-7b \\
& 4&  uukuguy&  speechless-llama2-13b&  databricks&  dolly-v2-12b \\
& 5&  TinyLlama&  TinyLlama-1.1B-Chat-v1.0&  h2oai&  h2o-danube-1.8b-chat \\
& 6&  HuggingFaceH4&  zephyr-7b-beta&  TinyLlama& TinyLlama-1.1B-Chat-v1.0 \\
& 7&  Deci&  DeciLM-7B-instruct&  HuggingFaceH4& zephyr-7b-beta \\
& 8&  Intel&  neural-chat-7b-v3-1&  jondurbin& bagel-8b-v1.0 \\
& 9&  hongzoh&  Yi-6B\_Open-Orca&  ikala& redpajama-3b-chat \\
& 10&  declare-lab&  flan-alpaca-xl&  HuggingFaceH4& zephyr-7b-alpha \\
& 11&  declare-lab&  flan-alpaca-xxl&  TheBloke& tulu-7B-fp16 \\
& 12&  declare-lab&  flan-alpaca-large&  TheBloke& tulu-13B-fp16 \\
& 13&  declare-lab&  flan-alpaca-base&  Deci& DeciLM-7B-instruct \\
& 14&  upstage&  SOLAR-10.7B-Instruct-v1.0&  garage-bAInd& Platypus2-7B \\
& 15&  openaccess-ai-collective&  jackalope-7b&  garage-bAInd& Platypus2-13B \\
& 16&  Open-Orca&  Mistral-7B-SlimOrca&  Open-Orca& Mistral-7B-OpenOrca \\
& 17&  luckychao&  TinyAlpaca-1.1B&  PygmalionAI& pygmalion-2-7b \\
& 18&  vilm&  Quyen-v0.1&  PygmalionAI& pygmalion-2-13b \\
& 19&  openaccess-ai-collective&  jackalope-7b&  uukuguy& speechless-llama2-13b \\
& 20&  M4-ai&  Orca-2.0-Tau-1.8B&  CalderaAI& 13B-Ouroboros \\
\cline{1-6}
\multicolumn{2}{l|}{\multirow{2}{*}{\textbf{Dataset}}} & \multicolumn{2}{c|}{\textbf{Slimorca}} & \multicolumn{2}{c}{\textbf{Openhermes}} \\
\cline{3-6}
\multicolumn{2}{l|}{} & Owner & Name&  Owner & Name \\
\cline{1-6}
\multirow{10}{*}{\rotatebox[origin=c]{90}{Member Models}}
& 1&   ajibawa-2023  & SlimOrca-13B & teknium & OpenHermes-2.5-Mistral-7B\\
& 2&   jondurbin & bagel-8b-v1.0 & NousResearch & Hermes-2-Theta-Llama-3-8B\\
& 3&   Deci & DeciLM-7B-instruct & NousResearch & Nous-Hermes-2-SOLAR-10.7B\\
& 4&   CallComply & DeciLM-7B-Instruct-128k & NousResearch & Hermes-2-Pro-Llama-3-8B\\
& 5&   Intel & neural-chat-7b-v3 & NousResearch & Hermes-2-Pro-Mistral-7B\\
& 6&   chargoddard & mistral-11b-slimorca & vilm & Quyen-Pro-v0.1\\
& 7&   jondurbin & bagel-7b-v0.1 & vilm & Quyen-SE-v0.1\\
& 8&   jondurbin & bagel-7b-v0.4 & vilm & Quyen-Plus-v0.1\\
& 9&   jondurbin & bagel-7b-v0.5 & vilm & Quyen-v0.1\\
& 10&   augmxnt & shisa-7b-v1 & vilm & Quyen-Pro-Max-v0.1\\
\cline{1-6}
\multirow{20}{*}{\rotatebox[origin=c]{90}{Non-Member Models}}
& 1&  databricks&  dolly-v2-3b&  databricks&  dolly-v2-3b \\
& 2&  databricks&  dolly-v2-7b&  databricks&  dolly-v2-7b \\
& 3&  databricks&  dolly-v2-12b&  databricks&  dolly-v2-12b \\
& 4&  SeaLLMs&  SeaLLM-7B-v2.5&  ikala&  bloom-zh-3b-chat \\
& 5&  shadowml&  BeagSake-7B&  ikala&  redpajama-3b-chat \\
& 6&  ikala&  redpajama-3b-chat&  HuggingFaceH4&  zephyr-7b-beta \\
& 7&  AtAndDev&  ShortKing-1.4b-v0.1&  HuggingFaceH4&  zephyr-7b-alpha \\
& 8&  malhajar&  meditron-7b-chat&  Intel&  neural-chat-7b-v3-1 \\
& 9&  HuggingFaceH4&  zephyr-7b-beta&  h2oai&  h2o-danube-1.8b-chat \\
& 10&  TinyLlama&  TinyLlama-1.1B-Chat-v1.0&  openaccess-ai-collective&  jackalope-7b \\
& 11&  HuggingFaceH4&  zephyr-7b-alpha&  CalderaAI&  13B-Ouroboros \\
& 12&  garage-bAInd&  Platypus2-13B&  uukuguy& speechless-llama2-13b \\
& 13&  Locutusque&  Orca-2-13b-SFT-v4&  TinyLlama& TinyLlama-1.1B-Chat-v1.0 \\
& 14&  Doctor-Shotgun&  TinyLlama-1.1B-32k-Instruct&  hongzoh& Yi-6B\_Open-Orca \\
& 15&  lgaalves&  gpt2\_camel\_physics-platypus&  luckychao& TinyAlpaca-1.1B \\
& 16&  allenai&  open-instruct-llama2-sharegpt-7b&  Open-Orca& Mistral-7B-OpenOrca \\
& 17&  allenai&  open-instruct-sharegpt-7b&  shadowml& BeagSake-7B \\
& 18&  openchat&  openchat-3.5-1210&  lgaalves& mistral-7b-platypus1k \\
& 19&  blueapple8259&  TinyAlpaca-v0.1&  AtAndDev& ShortKing-1.4b-v0.1 \\
& 20&  lgaalves&  mistral-7b-platypus1k&  malhajar& meditron-7b-chat \\
\cline{1-6}
\end{tabular}
}
\end{center}

\clearpage
\begin{table}[t]
\centering
\renewcommand{\arraystretch}{1.2}
\setlength\tabcolsep{21.4pt}
\footnotesize
\caption{\revise{Dataset inference results of \dia and our method under the non-IID setting in Databricks dataset. \texttt{HF} refers to models from Hugging Face. A value \texttt{x/y} indicates that \texttt{x}  models are correctly identified as members/non-members among \texttt{y} models.\label{tab:dia_inference}}}
\revise{
\begin{tabular}{l|c|c}
\toprule
\textbf{Models / Metrics}        & \textbf{\dia} & \textbf{Our Method}  \\
\toprule           
 Local Members            &10/10            &10/10              \\
\midrule
HF Members     &10/10            &10/10               \\
\midrule                          
HF Non-members &20/20            &20/20              \\
\midrule                           
Recall (\%)                & 100.0            & 100.0             \\
\midrule                        
Precision (\%)                   & 100.0            & 100.0            \\
\midrule
F1 (\%)                   & 100.0           &  100.0           \\
\bottomrule                     
\end{tabular}}
\end{table}

\begin{table}[t!]
        \footnotesize
        \setlength\tabcolsep{1.0pt}
	\centering
	\caption{Important notations used in this paper.\label{tab:notations}}
	\begin{tabular}{cl}
		\toprule
		\textbf{Notations} &\textbf{Descriptions} \\
		\midrule
		  $\mathcal{M}$ & The suspect model\\
            \midrule
            $\mathcal{D}$ & The victim dataset\\
            \midrule
		  $\mathbf{R}$ & The reference models never trained on $\mathcal{D}$\\
            \midrule
		  $\mathbf{R}^{t}$ & The reference models fine-tuned on $\mathcal{D}$\\ 
            \midrule
            $\mu$ & The length threshold used to pre-filter samples \\
            \midrule
            $\mathbf{R}(x)$ & The responses from $\mathbf{R}$ to samples passing pre-filtering\\
             \midrule
		  $\mathbf{R}^t(x)$ & The responses from $\mathbf{R}^{t}$ to samples passing pre-filtering\\
            \midrule
            $\delta^t$ & The similarity disparity threshold in identifying tainted samples\\
            \midrule
            $\delta^s$ & The similarity disparity threshold in dataset inference\\
		\bottomrule
	\end{tabular}%
\end{table}%
\begin{table}[t]
\centering
\renewcommand{\arraystretch}{1.2}
\setlength\tabcolsep{17.1pt}
\footnotesize
\caption{\revise{Dataset inference results of our method under the non-IID setting in Databricks dataset with varying value of $\delta^t$ and $\delta^s$. \texttt{HF} refers to models from Hugging Face. A value \texttt{x/y} indicates that \texttt{x}  models are correctly identified as members/non-members among \texttt{y} models.\label{tab:threshold}}}
\revise{
\begin{tabular}{l|c|c|c}
\toprule
\textbf{Models / Metrics}        & \textbf{0.1} & \textbf{0.2} & \textbf{0.3} \\
\toprule           
 Local Members            &3/10            &7/10        &10/10      \\
\midrule
HF Members     &4/10            &6/10          &10/10     \\
\midrule                          
HF Non-members &18/20            &18/20         &20/20     \\
\midrule                           
Recall (\%)                & 35.0            & 65.0         & 100.0    \\
\midrule                        
Precision (\%)                   & 77.8            & 86.7       & 100.0     \\
\midrule
F1 (\%)                   & 48.3           &  74.3      &  100.0     \\
\bottomrule                     
\end{tabular}}
\end{table}
\begin{table}[]
        \setlength\tabcolsep{37pt}
	\centering
        \footnotesize
	\caption{Local models used in our evaluations. \label{tab:local_models}}
	\begin{tabular}{ll}
		\toprule
		\textbf{Owner} & \textbf{Name} \\
        \toprule
	mistralai & Mistral-7B-v0.3 \cite{mistral7Bv3}\\
		google & gemma-2b \cite{gemma2B}\\
		meta-llama & Llama-3.1-8B \cite{llama3.1_8B}\\
        meta-llama & Llama-3.2-3B \cite{llama3.2_3B}\\
        Qwen & Qwen1.5-4B \cite{qwen}\\
        Qwen & Qwen1.5-7B \cite{qwen}\\
        Qwen & Qwen2.5-3B \cite{qwen2.5}\\
        Qwen & Qwen2.5-7B \cite{qwen2.5}\\
        THUDM & glm-2b \cite{DBLP:conf/acl/DuQLDQY022}\\
        THUDM & glm-10b \cite{DBLP:conf/acl/DuQLDQY022}\\
		\bottomrule
	\end{tabular}%
\end{table}%
\begin{table}[!t]
\centering
\renewcommand{\arraystretch}{1.2}
\footnotesize
\caption{Dataset inference results without selecting tainted samples (under non-IID setting). ``HF'' refers to Hugging Face. A value ``x/y'' indicates that ``x'' out of ``y'' members or non-members are correctly identified. \label{tab:eval:all_samples}}
\setlength\tabcolsep{5.5pt}
\begin{tabular}{l|c|c|c|c}
\toprule
\textbf{Dataset} & \textbf{Databricks} & \textbf{Alpaca} & \textbf{Slimorca} & \textbf{Openhermes} \\
\toprule
 Local Members            & 0/10           & 0/10       &  0/10        & 0/10           \\
 \cline{1-5}
 HF Members & 0/10           & 0/10       & 0/10         &  0/10          \\
 \cline{1-5}
 HF Non-members & 20/20           & 20/20       & 20/20         &  20/20          \\
 \cline{1-5}                        
 Precision (\%)                &  N/A          &  N/A      &   N/A       &    N/A        \\
 \cline{1-5}                        
 Recall (\%)                   &  0.0          & 0.0       &  0.0        & 0.0           \\
 \cline{1-5}
F1 (\%)                   &  N/A          & N/A       & N/A         & N/A           \\
\bottomrule                                            
\end{tabular}
\end{table}
\begin{table}[t]
\setlength\tabcolsep{4.3pt}
\centering
\caption{\revise{DPDLLM results with tainted samples under Non-IID and IID Settings. ``HF'' refers to Hugging Face. ``Non-members (1)'' represent local, non-member models that share the architectures of the reference models. ``Non-members (2)'' are wild, non-member models with architectures different from the reference models. A value ``x/y'' indicates that ``x'' out of ``y'' members or non-members are correctly identified.} }
\label{tab:baseline_tainted}
\renewcommand{\arraystretch}{1.2}
\revise{
\begin{tabular}{lcc}
\toprule
\textbf{Settings / Models} & \textbf{Without Tainted Samples} & \textbf{With Tainted Samples} \\
\midrule
\multicolumn{3}{l}{\textbf{Under Non-IID Setting}} \\ 
\midrule
Local Members      & 4/10  & 5/10  \\
HF Members         & 10/10 & 10/10 \\
HF Non-members     & 15/20 & 15/20 \\
\midrule
\multicolumn{3}{l}{\textbf{Under IID Setting}} \\
\midrule
Non-members (1)    & 3/5   & 4/5   \\
Non-members (2)    & 15/20 & 14/20 \\
\bottomrule
\end{tabular}}
\end{table}

\begin{table}[!t]
\centering
\renewcommand{\arraystretch}{1.2}
\footnotesize
\caption{Results of dataset inference for sequential training. ``D'' refers to Databricks, ``A'' refers to Alpaca, ``S'' refers to Slimorca, ``O'' refers to Openhermes.\label{tab:eval:seq_train}}
\setlength\tabcolsep{5.5pt}
\begin{tabular}{l|c|c|c|c}
\toprule
\textbf{Dataset} & \textbf{Databricks} & \textbf{Alpaca} & \textbf{Slimorca} & \textbf{Openhermes} \\
\toprule
 Mistral-7B-v0.3            & D           & D, A       &   A, S       &   A, S, O         \\
 \cline{1-5}
 gemma-2b                    & D           & D, A       &   A, S       &   S, O         \\
 \cline{1-5}
 Llama-3.1-8B               & D        & D, A       &  D, A, S     &   A, S, O       \\
 \cline{1-5}                        
 Llama-3.2-3B                &  D          &  D, A      &   A, S      &    S, O       \\
 \cline{1-5}                        
 Qwen1.5-4B                   &  D          & D, A       &  A, S       &   S, O       \\
 \cline{1-5}
 Qwen1.5-7B                   &  D          & D, A       &  A, S       &   S, O        \\
 \cline{1-5}
 Qwen2.5-3B                   &  D          & D, A       &  A, S       &   S, O        \\
 \cline{1-5}
 Qwen2.5-7B                   &  D          & D, A       &  D, A, S        &   S, O        \\
 \cline{1-5}
 glm-2b                   &  D          & D, A       &  A, S      &   S, O       \\
 \cline{1-5}
 glm-10b                   &  D          & D, A       &   D, A, S       &   A, S, O         \\
\bottomrule                                            
\end{tabular}
\end{table}
\begin{table}[]
        \setlength\tabcolsep{6pt}
	\centering
        \footnotesize
	\caption{Wild models used in the evaluation of extension to text generation domain. \label{tab:text_gen_models}}
	\begin{tabular}{c|c|c|c}
		\toprule
        \multicolumn{2}{c|}{Member Models} & \multicolumn{2}{c}{Non-member Models} \\
        \midrule
		\textbf{Owner} & \textbf{Name} & \textbf{Owner} & \textbf{Name}\\
        \midrule
	Alibaba-NLP & gte-large-en-v1.5 & declare-lab & flan-alpaca-large\\
		mosaicml & mpt-30b & databricks & dolly-v2-12b\\
		stanfordnlp & mrt5-small & HuggingFaceH4 & zephyr-7b-beta\\
        stanfordnlp & mrt5-large& PygmalionAI & pygmalion-2-13b\\
        anto18671 & lumenspark& garage-bAInd & Platypus2-13B\\
		\bottomrule
	\end{tabular}%
\end{table}%

\begin{table}[]
\centering
\renewcommand{\arraystretch}{1.2}
\footnotesize
\caption{Dataset inference results of text generation task. ``HF'' refers to Hugging Face. A value ``x/y'' indicates that ``x'' out of ``y'' members or non-members are correctly identified. \label{tab:eval:text_gen}}
\setlength\tabcolsep{42pt}
\begin{tabular}{l|c}
\toprule
 HF Members & 5/5           \\
 \midrule
 HF Non-members & 5/5          \\
 \midrule                        
 Precision (\%)                &  100.0 \\
 \midrule                        
 Recall (\%)                   &  100.0\\
 \midrule
F1 (\%)                   &  100.0 \\
\bottomrule                                            
\end{tabular}
\end{table}
\begin{table}[!t]
        \footnotesize
	\centering
	\caption{Time cost of our dataset inference method.\label{tab:efficiency}}
    \setlength\tabcolsep{14.5pt}
	\begin{tabular}{lccc}
		\toprule
		\textbf{Dataset} & \textbf{Offline} & \textbf{Online} & \textbf{Online Ratio} \\
		\toprule
		  Databricks &  4.0h&     0.25h &   6.25\%   \\
		 \midrule
		Alpaca&  11.5h&     0.25h &  2.17\%   \\
		 \midrule
		Slimorca& 18.0h&     0.25h &   1.39\%  \\
	\midrule
		Openhermes&  18.0h&     0.25h&     1.39\% \\
		\bottomrule
	\end{tabular}%
\end{table}%

\begin{figure}[t]
    \centering
    \begin{subfigure}{\columnwidth}
        \centering
        \includegraphics[width=\columnwidth]{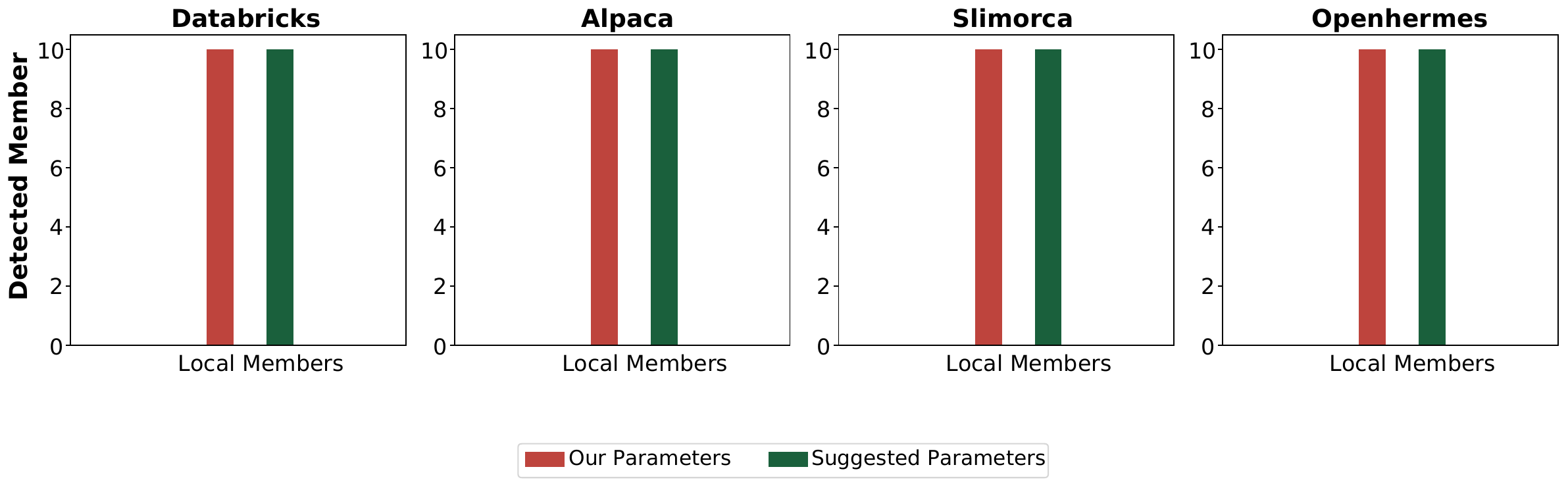}
        \caption{Suspect models trained with different parameters under non-IID}
        \label{fig:suspect_nonIID}
        \vspace{0.5em}
    \end{subfigure}
    \begin{subfigure}{\columnwidth}
        \centering
        \includegraphics[width=\columnwidth]{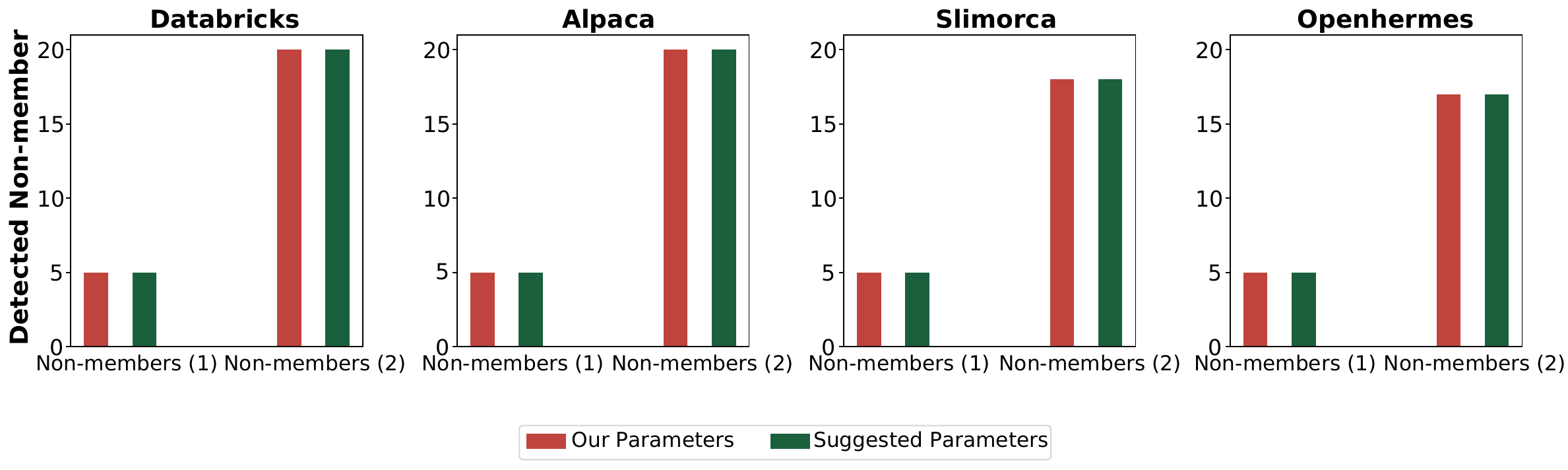}
        \caption{Suspect models trained with different parameters under IID}
        \label{fig:suspect_IID}
        \vspace{0.5em}
    \end{subfigure}
    \caption{Robustness of our methods against different values of hypeparameters for suspect models. ``Non-members (1)'' represent local, non-member models that share the architectures of the reference models. ``Non-members (2)'' are wild, non-member models with architectures different from the reference models.}
    \label{fig:suspect_para}
\end{figure}

\begin{figure}[t]
    \centering
    \includegraphics[width=0.475\textwidth]{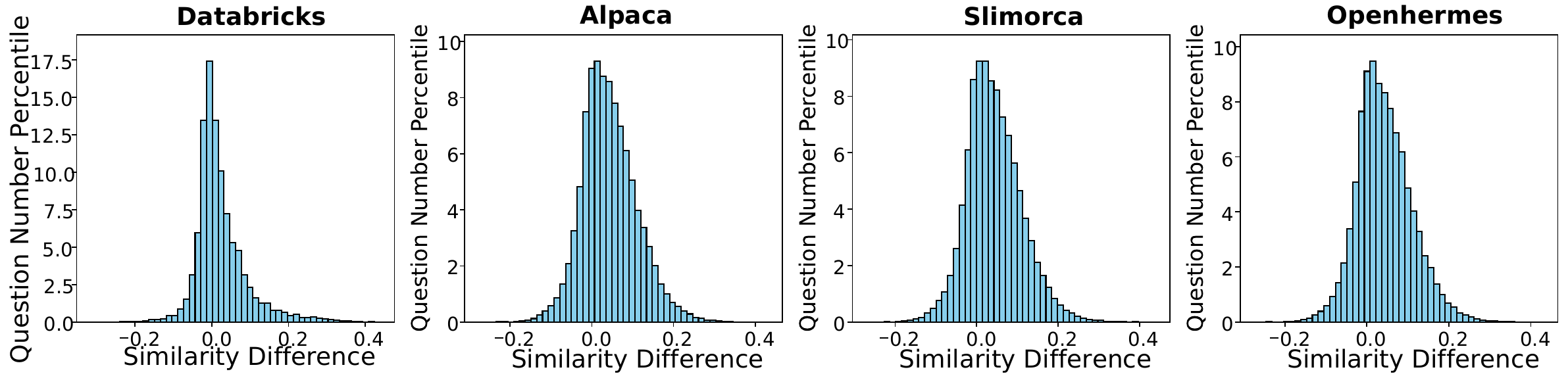}
    \caption{Distribution of similarity difference between the suspect model and the reference models (before and after fine-tuning) on all samples from the target dataset. Given a sample, we calculate a similarity score between the suspect model's response and the non-fine-tuned reference models' responses. Likewise, we calculate another similarity score using the fine-tuned reference models. The difference between the two scores for all samples is used to plot the distribution.}
    \label{fig:all-sample-hist}
\end{figure}

\begin{figure}[htbp]
    \centering
    \includegraphics[width=0.475\textwidth]{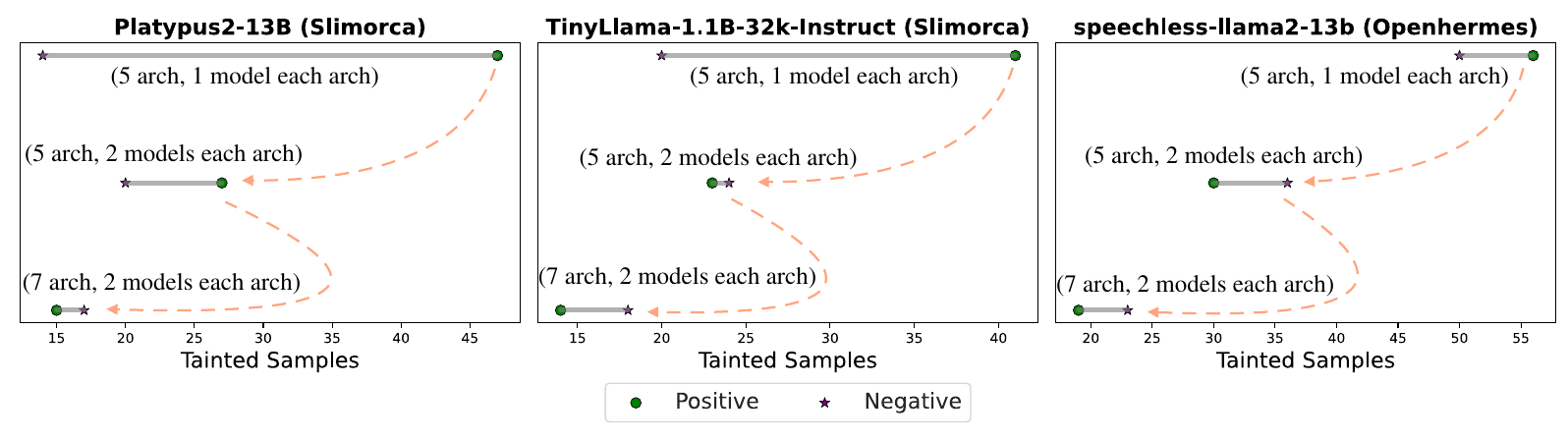}
    \caption{Impacts of reference model number and architecture diversity on the distribution of tainted samples under IID settings.}
    \label{fig:ref-tainted-change}
\end{figure}

\begin{figure}[htbp]
    \centering
    \begin{subfigure}{\columnwidth}
        \centering
        \includegraphics[width=\columnwidth]{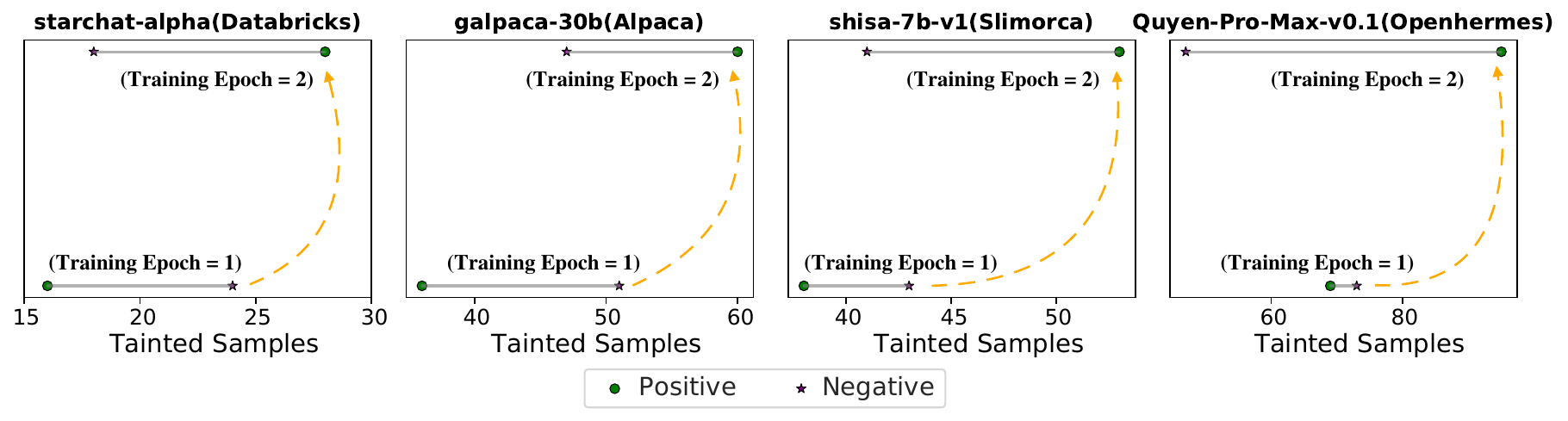}
        \caption{Distribution of tainted samples under Non-IID settings.}
        \label{fig:ref_epoch_tainted_noniid}
        \vspace{0.5em}
    \end{subfigure}
    \begin{subfigure}{\columnwidth}
        \centering
        \includegraphics[width=\columnwidth]{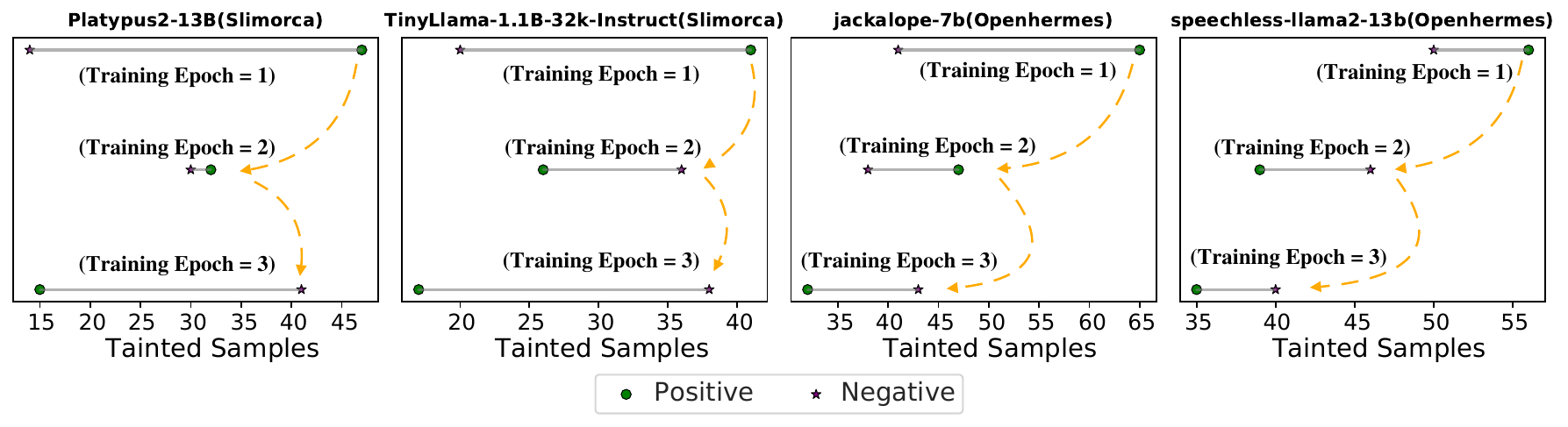}
        \caption{Distribution of tainted samples under IID settings.}
        \label{fig:ref_epoch_tainted_iid}
        \vspace{0.5em}
    \end{subfigure}
    \caption{Impacts of training epoch for reference model on the distribution of tainted samples.}
    \label{fig:ref_epoch_tainted}
\end{figure}

\begin{figure}[htbp]
    \centering
    \includegraphics[width=0.475\textwidth]{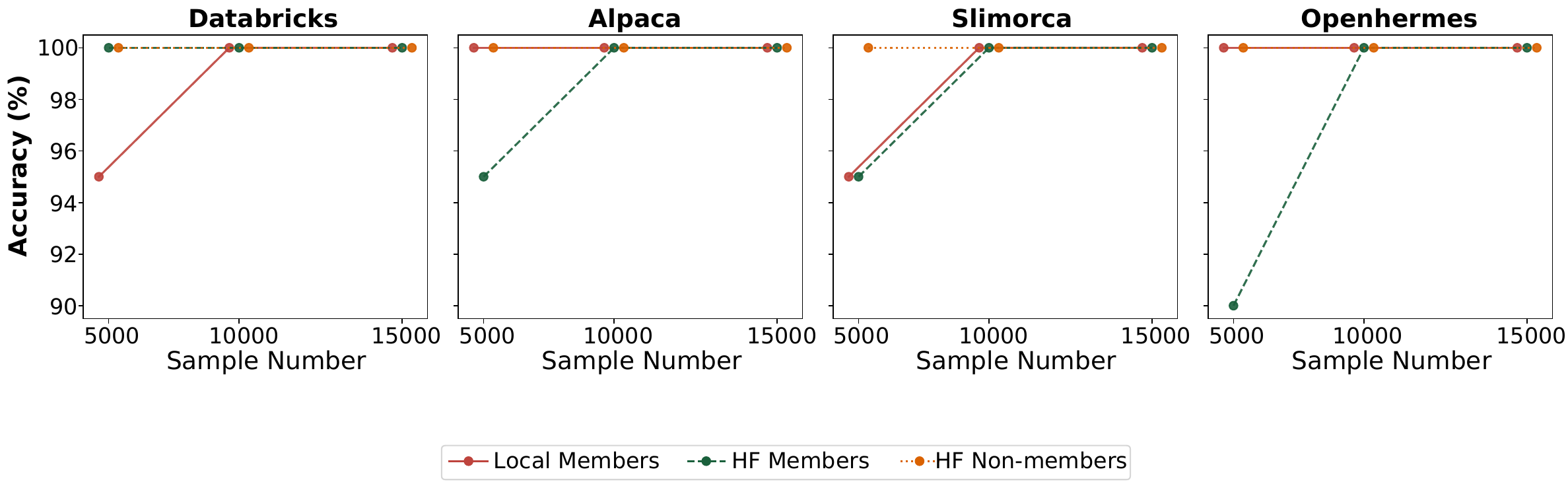}
    \caption{Results of dataset inference using different sizes of the original datasets.  ``HF'' refers to Hugging Face.}
    \label{fig:dataset_trend}
\end{figure}

\begin{figure}[htbp]
    \centering
    \includegraphics[width=0.475\textwidth]{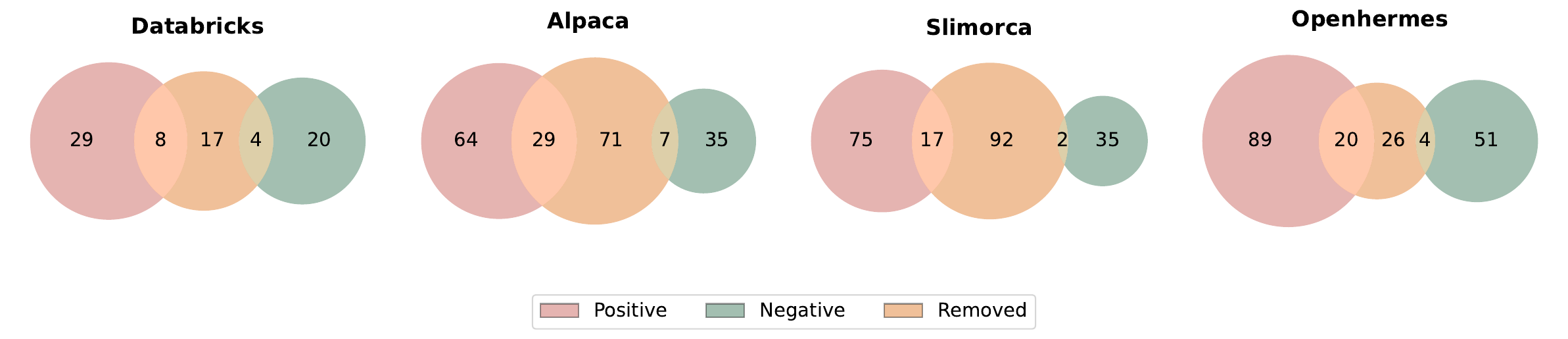}
    \caption{Overlap of tainted samples for Llama-3.2-3B identified by our method against evasion of removing tainted samples.}
    \label{fig:tainted-overlap}
\end{figure}

\begin{figure}[htbp]
    \centering
    \includegraphics[width=0.475\textwidth]{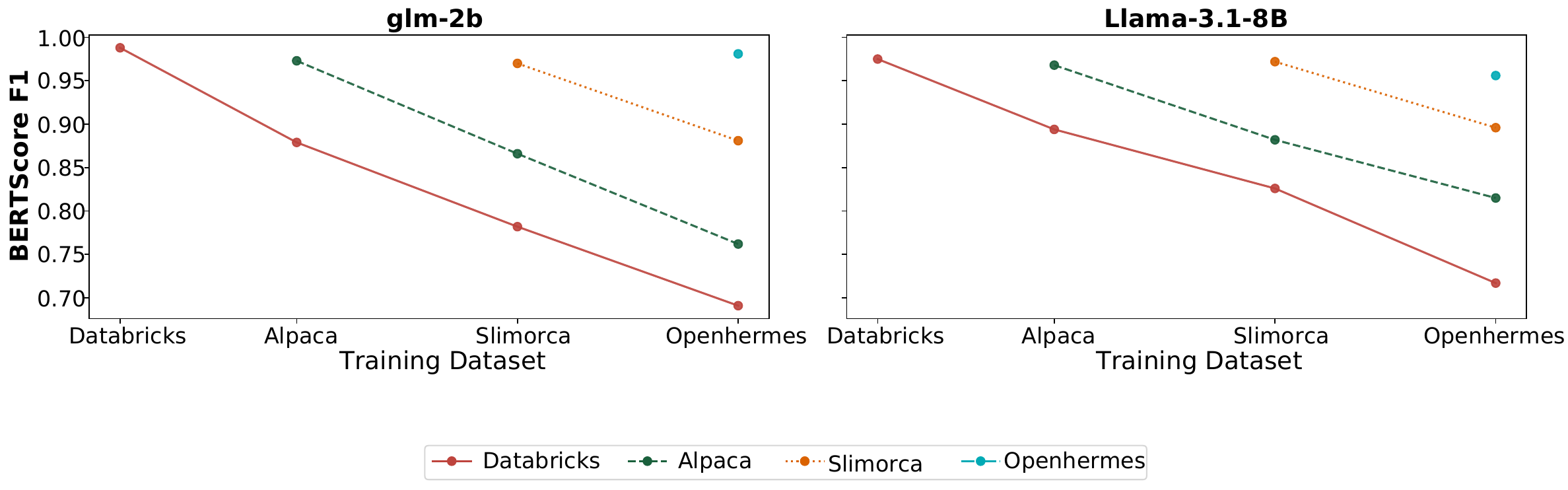}
    \caption{Utility performance of models on datasets in sequential manner.}
    \label{fig:utility_trend}
\end{figure}

\end{document}